\documentclass[10pt,
 onecolumn,
 amsmath,amssymb,
 aps, physrev,
]{revtex4-2}
\usepackage{bm}
\usepackage{lipsum} 
\usepackage{xcolor}
\usepackage{amsbsy}
\usepackage{lineno}
\usepackage{tikz}
\usepackage{float}
\usepackage{placeins}
\usepackage{footnote}
\usepackage{microtype}  
\usepackage{multirow}
\usepackage{eqnarray, mathrsfs}
\usepackage{subfigure}
\usepackage{colortbl}
\usepackage{xcolor}
\usepackage{footnote}
\usepackage[flushleft]{threeparttable}
\usepackage{graphicx}
\usepackage{dcolumn}
\usepackage{bm}
\usepackage{array}
\usepackage[colorlinks, linkcolor=black,citecolor=blue,urlcolor=blue]{hyperref}
\usepackage{booktabs}  
\usepackage{placeins}
\newcommand{\note}[1]{\textcolor{blue}{#1}}
\DeclareMathOperator\arctanh{arctanh}

\begin{document}

\title{Fitness Inference in Presence of Migrations between Coupled Evolving Populations}%

\author{Yu-Han Huang}
\thanks{These authors contributed equally to this work.}
\affiliation{ 
School of Science, Nanjing University of Posts and Telecommunications, 210023, CHINA
}%
\author{Bastien Dumont}%
\thanks{These authors contributed equally to this work.}
\affiliation{Department of Physics, Université Paris-Saclay, FRANCE  
}%

\author{Hong-Li Zeng}
\email{hlzeng@njupt.edu.cn}
\affiliation{School of Science, Nanjing University of Posts and Telecommunications, \\ National Laboratory of Solid State Microstructures, Nanjing University, Nanjing 210093, CHINA; \\ Key Laboratory of Radio and Micro-Nano Electronics of Jiangsu Province, Nanjing, 210023, CHINA}

\author{John Barton}%
\email{jpbarton@pitt.edu}
\affiliation{%
Department of Computational \& Systems Biology, University of Pittsburgh School of Medicine, USA
}%
\author{Erik Aurell}  
\email{eaurell@kth.se}
\affiliation{
Department of Computational Science and Technology, AlbaNova University Center, SE-106 91 Stockholm, SWEDEN\\} 
\date{\today}

\begin{abstract}
The phase of Quasi-Linkage Equilibrium (QLE) in evolutionary populations is analogous to the thermal equilibrium state in statistical mechanics, a concept pioneered by Kimura in 1965 for two-locus two-allele models. QLE describes a stationary state maintained by the interplay of selection, mutation, recombination and genetic drift. Here we extend QLE theory to populations connected by migration, a fundamental evolutionary force that couples the evolutionary dynamics of interacting subpopulations. Specifically, we examine two populations interacting via symmetric or asymmetric migration while evolving under multi-locus selection. Using whole-genome time-series data generated through FFPopSim, we demonstrate that the QLE phase persists under conditions of sufficiently low migration rates. In this regime, we derive analytical inference relations that allow for the accurate and quantitative estimation of both additive fitness and epistatic interactions. 

\begin{description}
    \item[keywords] 
 QLE $|$  Migration $|$ Epistasis $|$ Selection Coefficients
\end{description}
\end{abstract}

\maketitle

\section{Introduction} \label{sec:Introduction}

Understanding the patterns of genetic variation in natural populations and the underlying evolutionary mechanisms that shape them is a central goal in population genetics \cite{kimura1965attainment}. Over the past few decades, the rapid development of genomic sequencing technologies has enabled researchers to analyze the genetic structure of populations at a genome-wide scale. This advancement has provided unprecedented opportunities to uncover the genetic basis of complex adaptive traits \cite{neher2009competition}. However, extracting information about selection pressures, epistatic interactions, and population structure from sequence data remains a significant challenge in computational biology.

The theory of Quasi-Linkage Equilibrium (QLE) provides an important theoretical framework for addressing this problem. First proposed by Kimura in 1965 \cite{kimura1965attainment} and later extended to a genome-wide scale by Neher and Shraiman \cite{NeherShraiman2011,zeng2020inferring, weinreich2005perspective}, QLE theory predicts that under high recombination rates, the distribution of genotypes in a population approaches a form analogous to a thermal equilibrium state in statistical physics. In this phase, the genotype probability distribution can be described by an exponential family distribution similar to the Ising model, making it possible to infer the underlying fitness landscape from population data \cite{de2014empirical}.

The core of QLE theory lies in establishing a quantitative relationship between observable patterns of genetic variation (such as allele frequencies and linkage disequilibrium) and the underlying fitness parameters (including additive effects and epistatic interactions). This theoretical framework has been successfully applied in several fields, including protein structure prediction \cite{cocco2018inverse}, RNA interaction identification \cite{wright1943isolation}, and microbial genome analysis \cite{slatkin1985gene}. Notably, Zeng and Aurell \cite{de2014empirical} demonstrated through computer simulations that under appropriate parameter conditions, the inverse Ising/Potts method can be used to accurately infer epistatic fitness parameters.

However, classical QLE theory primarily focuses on a single, randomly mating population, often neglecting the spatial structure and migration dynamics that are common in natural populations \cite{li2024improved}. In the real world, populations are frequently divided into subpopulations, and migration of individuals between different habitats is a crucial driver of evolution \cite{mc1999indirect}. Migration not only alters local allele frequencies but may also influence the strength and patterns of gene interactions, thereby profoundly affecting the fitness landscape of the entire population. Therefore, extending QLE theory to incorporate migration processes is essential for understanding the evolutionary dynamics of structured populations \cite{rousset2004genetic}.

Migration has long been recognized as a central force in evolution because it couples populations that may otherwise experience distinct demographic histories and selective environments \cite{slatkin1985gene, lenormand2002gene}. Gene flow can reduce differentiation between subpopulations, but it can also introduce genetic variation and oppose or reshape local adaptation. This tension is especially important when selection varies across space: a mutation that is beneficial in one environment may be deleterious in another, so its observed trajectory depends jointly on local fitness effects, migration rates, population sizes, and sampling location.

Early efforts to represent migration mathematically replaced a single, well-mixed population with separated but interacting subpopulations. Wright’s island model and his work on isolation by distance provided a foundation for asking how migration and drift determine differentiation among subpopulations \cite{wright1931evolution, wright1943isolation}. Kimura and Weiss’s stepping-stone model made spatial locality explicit, showing how genetic correlation decays with geographic distance when migration occurs mainly between neighboring subpopulations \cite{kimura1964stepping}. In parallel, migration-selection models asked when spatially varying selection can maintain allele-frequency differences despite gene flow. Haldane’s cline theory related cline shape to the balance between dispersal and selection, and later work on clines developed this idea into a framework for estimating the strength of barriers, dispersal, and selection from spatial allele-frequency patterns \cite{haldane1948theory, slatkin1973gene, barton1979gene}.

Some later theoretical developments have focused on genealogies and statistical inference. The structured coalescent considered evolution among subpopulations with migration \cite{notohara1990coalescent}, which enabled likelihood-based inference of migration rates, effective population sizes, and divergence times from genetic data \cite{beerli2001maximum, nielsen2001distinguishing}. Subsequent studies have built on these ideas to identify loci under selection in populations with spatial structure \cite{coop2010using, gunther2013robust, frichot2013testing}, however, comparatively few approaches have been developed to directly estimate the fitness effects of mutations in these contexts \cite{mathieson2013estimating, lyu2022inferring}.

In this work, building on the work of Zeng and Aurell \cite{de2014empirical}, we develop an extended QLE framework that incorporates migration. We consider two subpopulations interacting through migration, with each subpopulation subject to selection, mutation, and recombination. In this setting, we derive modified QLE relations in the presence of migration, develop a corresponding algorithm for fitness inference, and identify conditions for accurate inference through systematic parameter exploration in simulations.

We find that the introduction of migration significantly alters the characteristics of the QLE phase. When correcting for migration, we find that fitness inference is generally accurate, especially in favorable parameter regimes (i.e., moderate mutation rate, moderate recombination rate, low migration rate, moderate additive effect strength ($\sigma_1$), and low epistatic effect strength ($\sigma_2$)). These findings not only validate the effectiveness of the extended theory but also provide practical guidelines for analyzing genomic data from real, structured populations.

\section{Quasi-Linkage Equilibrium in the presence of migration}

We model the migration of individuals between two finite subpopulations A and B following the Moran model for genetics~\cite{moran1958random}. This is built on discrete time intervals, where only one event can take place 
at the same interval. In the Moran model either an individual dies and a new one appears, or an individual mutates. 
We will consider each individual to be represented by a genome of fixed length $L$, with the allele at each locus $i$ taking on one of two possible states $s_i\in\{-1, 1\}$. 
Here, we will neglect more complex genetic alterations such as insertions and deletions which change the length of the genome (number of genetic loci). 

The natural evolutionary forces in population genetics include selection, mutation, recombination, and genetic drift. Selection is the fundamental driving force of evolution 
determining the probability that an individual passes its genetic material to the next generation. Here, selection acts through a fitness function that assigns a fitness value $F(\mathbf{g})$ to each genotype $\mathbf{g}=(s_1,s_2,\ldots,s_L)$,
\begin{equation}
\label{eq:F-definition}
    F(\mathbf{g}) = f_0 + \sum_i f_i s_i
    +\sum_{i<j} f_{ij}s_is_j\,.
\end{equation}
$F(\mathbf{g})$ quantifies natural selection and hence the likely future relative abundance of genotype $\mathbf{g}$. The coefficients 
$f_i$ are known as \textit{additive fitness} and the
coefficients $f_{ij}$ are \textit{epistatic fitness}
by \textit{pairwise effects}.
We here consider the problem on inferring additive and pairwise epistatic fitness; higher-order epistatic fitness effects are not considered.
The constant $f_0$ plays no role in that evolution and will hence from now on be set to zero.

Classical theories typically assume that populations undergo panmictic mating, neglecting the influences of spatial structure and individual migration. Recent studies have formally recognized gene flow as the ``fifth evolutionary force", demonstrating that low migration rates can maintain signals of local selection \cite{battey2020space}, while high migration rates can obscure polygenic adaptive landscapes. Ringbauer \textit{et al.} \cite{ringbauer2021parental} further pointed out that migration reshapes the linkage disequilibrium landscape at the genomic scale, necessitating corrections to the original QLE equations. We will here derive the cumulant dynamics incorporating migration within the QLE framework, providing closed-form analytical expressions for extending the Kimura–Neher–Shraiman (KNS) theory. Building 
on our previous work reviewed in 
\cite{dichio2023statistical}
and recently carried forward in
\cite{Zeng2026}, as well as other
foundational work \cite{schraiber2016bayesian,mavroudi2023global}, the present study incorporates migration into the QLE framework, constructing a model with migration between two subpopulations. We find that migration introduces coupled correction terms in the cumulant dynamical equations, thereby altering the inference method and accuracy of fitness parameters.

\subsection{The Gibbs distribution of the QLE state}
Quasi-Linkage Equilibrium was first introduced by Kimura in 1965 in a bi-allelic 2-locus model \cite{kimura1965attainment}. It was later developed by Neher and Shraiman in a genome-wide setting~\cite{neher2009competition,NeherShraiman2011}. In those cases, the recombination rate was high, so while alleles at different loci remain statistically dependent, they are only weakly correlated. The QLE phase exhibits many similarities to a thermal equilibrium state. 
A formal definition posed in \cite{dichio2023statistical} 
is as follows: a population is said to be in a Quasi-Linkage Equilibrium phase if the two following conditions are met:
\begin{itemize}
    \item multi-genome distributions approximately factorize \textit{i.e.}
    $P_k(g_1,g_2,\dots,g_k,t) = P(g_1,t) P(g_2,t) \dots P(g_k,t)$.
    \item single-genome distributions lie in an exponential family with terms of order no higher than those appearing in the fitness function.
\end{itemize}
For a fitness function which has, at most, pairwise epistatic interactions and where all loci are bi-allelic, 
as in \eqref{eq:F-definition}, the second condition implies that the distribution of individuals over genotypes is the Gibbs distribution of an Ising model:
\begin{equation}
P(g,t) = \frac{1}{\mathcal{Z}(t)} \exp\left( \sum_{i=1}^{L} h_{i}(t) s_{i} + \sum_{i<j} J_{ij}(t) s_{i} s_{j} \right)
\label{eq:QLE}
\end{equation}
In contrast to thermal equilibrium, the coefficients $h_i$ and $J_{ij}$ are here not necessarily time-independent, and are not directly interpretable as physical energies or biological fitness effects. The quantitative relation between fitness as
in \eqref{eq:F-definition} and distribution parameters
in \eqref{eq:QLE} is the central content of the QLE theory.

Relation~\eqref{eq:QLE} is mathematically equivalent to the Gibbs distribution of an Ising model in statistical physics. Like other maximum-entropy models, it is fully characterized by a set of sufficient statistics, here the magnetizations $\chi_i$ and pairwise correlations $\chi_{ij}$. 
These are defined as
\begin{equation}
\chi_i = \langle s_i \rangle .
\label{eq:first order}
\end{equation}
and
\begin{equation}
\chi_{ij} = \langle s_i s_j \rangle - \langle s_i \rangle \langle s_j \rangle = \langle s_i s_j \rangle - \chi_i \chi_j.
\label{eq:second order}
\end{equation}
where $\left<\cdots\right>$ means averaging with respect 
to~\eqref{eq:QLE}. The relationship between on the one hand $(\chi_i,\chi_{ij})$ and on the other $(h_i,J_{ij})$ is an inverse statistical mechanics of Direct Coupling Analysis (DCA) problem which is intractable if done exactly for large systems, but for which a series of efficient approximations have been developed and widely studied in the literature~\cite{Nguyen2017}.

\subsection{The Kimura-Neher-Shraiman Theory of the QLE state}\label{sec:Presentation of the Kimura-Neher-Shriman Theory}
In the Kimura-Neher-Shraiman (KNS) formulation of QLE theory, pairwise Ising couplings are related to epistatic fitness coefficients through
\begin{equation}
J_{ij} = \frac{f_{ij}}{r  c_{ij}}
\label{eq:KNS}
\end{equation}
where $r$ is the recombination rate and $c_{ij}$ is the probability that loci $i$ and $j$ are inherited from different parents. For sufficiently distant loci under strong recombination, $c_{ij}$ approaches $1/2$. Eq.~\eqref{eq:KNS} therefore provides a direct inference relation for pairwise epistatic fitness effects:
\begin{equation} 
f_{ij}^* = \frac{1}{2}\, J_{ij}^* \cdot r 
\label{eq:epistatic_ﬁtness}
\end{equation}
where $*$ indicates the inferred value. $J_{ij}^*$ is determined from data, and the remaining parameter $r$ is an effective proportionality term. In practice, $r$ must be estimated independently. In the presence of mutations, the key relation becomes \cite{zeng2021inferring} 
\begin{equation}
J_{ij} = \frac{f_{ij}}{4\mu +r c_{ij}}.
\label{eq:KNS_mutation}
\end{equation}
For the parameters $h_i$ in the distribution in~\eqref{eq:QLE},
we get in the naive mean-field
version of DCA (hence, as an approximation, on the level of inverse statistical mechanics)
\begin{equation}
h_i = \arctanh \left(\chi_i\right) - \sum_{j} J_{ij} \chi_j ,
\label{eq:inverse_mean_field} 
\end{equation}
where $h_i$ are the parameters of the distribution in~\eqref{eq:QLE}, $\chi_j$ is the magnetization (or equivalently, allele frequency imbalance) at locus $j$. 
Neglecting mutation and migration, the corresponding additive inference relation is:
\begin{equation}
f_i^* = \dot{h}^*_i - \sum_{j} f_{ij}^* \, \chi_j\,.
\label{eq:Inferred_additive_fitness}
\end{equation}

\subsection{Extension of QLE to include migration}
We now extend the KNS framework to two interacting subpopulations (or ``islands''), denoted $A$ and $B$, which are connected by migration.
The genotype distribution in population $A$ evolves according to the master equation
\begin{equation}
\frac{d}{dt} P_A(g,t) = 
\frac{d}{dt}\Big|_{\text{fit}} P_A(g,t) + 
\frac{d}{dt}\Big|_{\text{mut}} P_A(g,t) + 
\frac{d}{dt}\Big|_{\text{rec}} P_A(g,t) + 
\frac{d}{dt}\Big|_{\text{mig}} P_A(g,t)
\label{eq:Modified_master_equation}
\end{equation}
where terms related to fitness, mutation and recombination remain unchanged compared to the original equation
presented in \cite{NeherShraiman2011}
and reviewed more recently in \cite{dichio2023statistical}.
Naturally, population $B$ follows the same equation. To derive the migration term, we first define the probability that an individual has the genome $g$ in the population $A$ by
\begin{equation}
     P_{A}(g,t)=\frac{n_{A}(g,t)}{N_{A}(t)},
\end{equation}
where $n_{A}(g,t)$ is the number of individuals with genome $g$ in the population $A$ at time $t$ and $N_{A}(t)$ is the size of subpopulation $A$. We assume sufficiently large population sizes that genotype frequencies evolve smoothly in time.
The migration rates between the two islands are $m_{A\to B}$ and $m_{B\to A}$. Assuming that the number of individuals exchanged during migration remains small compared to the total population size over a small time step $dt$ ($m\,dt \ll 1 $), calculations detailed in Supplementary Information yield at first order in $dt$: 
\[P_{A}(g, t + dt)= P_{A}(g, t) + m_{B\to A}\frac{N_{B}(t)}{N_{A}(t)}\left[P_{B}(g,t)-P_{A}(g,t)\right] dt.\]
The evolution of $P_{B}(g, t)$ is analogous, with the roles of $A$ and $B$ exchanged.
In what follows the time dependence $t$ will be dropped in order to lighten the notation. 
The time derivative of the probability of genome $g$ in the population $A$ due to migration is then
\begin{equation}
    \label{eq:mig}
    \left. \frac{d}{dt} \right|_{\text{mig}}P_{A}(g) =m_{B\to A}\frac{N_{B}}{N_{A}}\left[P_{B}(g)-P_{A}(g)\right].
\end{equation}
The migration term therefore drives the genotype distribution in population $A$ toward that of population $B$ (and vice versa), with a strength proportional to the incoming migration flux.

Next, we aim to derive a formula linking the fitness parameters ($f_{i}$, $f_{ij}$) with the Ising parameters ($h_{i}$, $J_{ij}$)
in the presence of migration.
The starting point of the inference is similar
to previous investigations of QLE~\cite{NeherShraiman2011,dichio2023statistical}. For simplicity we write it only for population $A$; equations for population $B$ are obtained by exchanging indices $A \leftrightarrow B$.
\begin{equation}
\frac{d}{dt}\ln P_{A}(g)=-\frac{\dot{\mathcal{Z}}_{A}}{\mathcal{Z}_{A}}+\sum_{i=1}^{L}\dot{h}_{i}^As_{i}+\sum_{i<j}\dot{J}_{ij}^As_{i}s_{j}. 
\label{eq:partition_function_log_deriv} 
\end{equation}
We then substitute on the left hand side the derivative of probability with respect to time due to selection, mutation, recombination and migration
and relate them to terms on the right hand side as to the spin monomials involved (constant, linear, quadratic). Matching coefficients of constant, linear, and quadratic spin monomials then yields the cumulant dynamics.
Detailed derivations are provided in the Supplementary Information.

The resulting inference relations for additive and epistatic fitness in population $A$ are
\begin{equation}
{f}_i^{A,*} = \dot{h}_i^A  + 2\mu h_i^A - r\sum_{j}c_{ij} J_{ij}^A \chi_j^A + 
m_{B\to A}
\frac{N_B}{N_A} \frac{\mathcal{Z}_A}{\mathcal{Z}_B} (h_i^A - h_i^B)
\label{eq:fi-inference-formula-v1}
\end{equation}
\begin{equation}
f_{ij}^{A,*} =  (4\mu +r c_{ij}) J_{ij}^A + m_{B\to A}\frac{N_B}{N_A}\frac{\mathcal{Z}_A}{\mathcal{Z}_B}
 \left( J_{ij}^A - J_{ij}^B \right).
\label{eq:fij-inference-formula-v1}
\end{equation}
Assuming that the population sizes $N_A$ and $N_B$ have reached a stationary migration balance, we have $N_B/N_A=m_{A\to B}/m_{B\to A}$. This yields
\begin{gather}
    \hspace{-0.1\textwidth}f_{i}^{A,*}=\dot{h}_i^A + 2\mu h_i^A-r\sum_{j\neq i}c_{ij}J_{ij}^A\chi_j^A+\left(h_i^A-h_i^B\right)\tilde{m}_{A\to B}
\label{eq:Inferred_additive_pop_A}\\[-0.08\baselineskip]
\hspace{0.6\textwidth}\text{(at equilibrium $A\leftrightarrow B$)}\nonumber\\[-0.08\baselineskip]
    \hspace{-0.1\textwidth}f_{ij}^{A,*} = J_{ij}^A(4\mu +r c_{ij}) + (J_{ij}^A - J_{ij}^B) \tilde{m}_{A\to B}
\label{eq:Inferred_epistasis_pop_A}
\end{gather}
with $\tilde{m}_{A\to B}=m_{A\to B}\frac{\mathcal{Z}_A}{\mathcal{Z}_B}$ 
and analogously for population $B$. 
Migration therefore enters into the inference equations through an effective migration rate that is scaled by the ratio of the partition functions of the two subpopulations.
The modified migration rate comes from the expression in \eqref{eq:mig} after inserting the Gibbs distribution ansatz (see Supplementary Information). Migration thus modifies the standard KNS inference relations (i.e., $f_{ij}^{KNS,*} = J_{ij}(4\mu+rc_{ij})$) through coupling terms between the two subpopulations.

We show here that it is possible to obtain two QLE phases in subpopulations evolving on their own and interacting through migration. Combining \eqref{eq:Inferred_epistasis_pop_A} and its equivalent for population B yields
\begin{equation}
\label{eq:inference_epistatis_mig}
    J_{ij}^A = \frac{1}{2(4\mu +r c_{ij})} \left[ 
    \left(f_{ij}^A + f_{ij}^B\right) 
    + \left(f_{ij}^A - f_{ij}^B\right)\frac{ 
    4\mu +r c_{ij} + \tilde{m}_{B\to A} - \tilde{m}_{A\to B}}
    {4\mu +r c_{ij} +  \tilde{m}_{B\to A} +\tilde{m}_{A\to B}} 
    \right]
\qquad\hbox{(at equilibrium $A\leftrightarrow B$)}
\end{equation}
The first term reflects the contribution of the mean epistatic fitness across both subpopulations, while the second captures the effect of their difference, weighted by the strength of the asymmetric migration relative to the other biological forces. 
If $f_{ij}^A=f_{ij}^B=f_{ij}$:
\begin{equation*}
J_{ij}^A = \frac{f_{ij}}{4\mu +r c_{ij}} = J_{ij} = J_{ij}^B.
\label{eq:J_{ij}^A = J_{ij}^B}
\end{equation*}
Thus, if the underlying epistatic fitness coefficients are identical in the two subpopulations, then the inferred couplings also become identical and the two subpopulations effectively behave as a single QLE system. 
In general, we may combine \eqref{eq:Inferred_epistasis_pop_A} and the analogous expression for population $B$ to obtain a compact expression for the difference of inferred epistatic fitness between the two populations
\begin{equation}
\Delta f_{ij}^{AB,*} = 
f_{ij}^{A,*}-f_{ij}^{B,*}
= \left(J_{ij}^A - J_{ij}^B\right)
\left(4\mu +r c_{ij} + \tilde{m}_{A\to B}
+\tilde{m}_{B\to A}\right)
\qquad\hbox{(at equilibrium $A\leftrightarrow B$)}
\label{eq:Inferred_epistasis_difference}
\end{equation}
In the case of symmetric migration, $m_{B\to A}=m_{A\to B}=m$, \eqref{eq:Inferred_epistasis_difference} can be simplified to
\begin{equation}
\Delta f_{ij}^{AB,*} 
= \left(J_{ij}^A - J_{ij}^B\right)
\left[4\mu +r c_{ij} + m\left(\frac{\mathcal{Z}_A}{\mathcal{Z}_B}+\frac{\mathcal{Z}_B}{\mathcal{Z}_A}\right)\right]
\qquad\hbox{(at equilibrium $A\leftrightarrow B$ and symmetric rates)}
\label{eq:Inferred_epistasis_difference_sym}
\end{equation}
Using a mean field approximation for the ratio of partition functions (detailed in the Supplementary Information), we have
\begin{equation*}
\frac{\mathcal{Z}_A}{\mathcal{Z}_B} 
= e^{-\left(F_A-F_B\right)}=e^{-\Delta F_{AB}}
\qquad 
\Delta F_{AB} = 
\frac{1}{2}\left[
\sum_{i=1}^L \log\frac{1-{\chi_i^A}^2}{1-{\chi_i^B}^2}
+ {\boldsymbol{\chi}^A}^\top J^A \boldsymbol{\chi}^A 
- {\boldsymbol{\chi}^B}^\top J^B \boldsymbol{\chi}^B
\right]
\label{eq:fraction_partition}
\end{equation*}
where $\Delta F_{AB}$ is the free energy difference between the two populations in a mean field approximation. Substituting this into \eqref{eq:Inferred_epistasis_difference_sym} yields:
\begin{equation}
\Delta f_{ij}^{AB,*}
= \left(J_{ij}^A-J_{ij}^B\right)
\left[4\mu +r c_{ij} + 2m\cosh\!\left(\Delta F_{AB}\right)\right].
\label{eq:fij_beta_final}
\end{equation}
Hence the inferred epistatic fitness parameters (left hand side) is proportional to the difference of Ising model parameters of QLE theory (first term on right hand side). The proportionality positively depends on the rates of recombination, mutation, and migration. On the other hand, the difference of Ising model parameters is a consequence of the true underlying epistatic fitness parameters being different, as shown by the forward relation \eqref{eq:inference_epistatis_mig}. The two relations therefore combine in a coherent scheme showing that two populations in QLE with migration between them can coexist in equilibrium, with migration only modifying the parameters describing each of the states.

\section{Simulation design for evolutionary populations under migration and QLE}
We used FFPopSim~\cite{zanini2012ffpopsim} simulations to test our migration-aware inference framework in structured populations evolving under selection, mutation, recombination, and migration.
We considered two subpopulations, denoted as $A$ and $B$, connected by bidirectional migration. 
Since the underlying fitness parameters are explicitly specified in the simulations, the inferred fitness parameters can be quantitatively compared to the true values. All simulations were conducted in high-recombination regimes intended to favor QLE-like states.

\begin{figure}[ht!]
    \centering
    \includegraphics[width=0.9\linewidth]{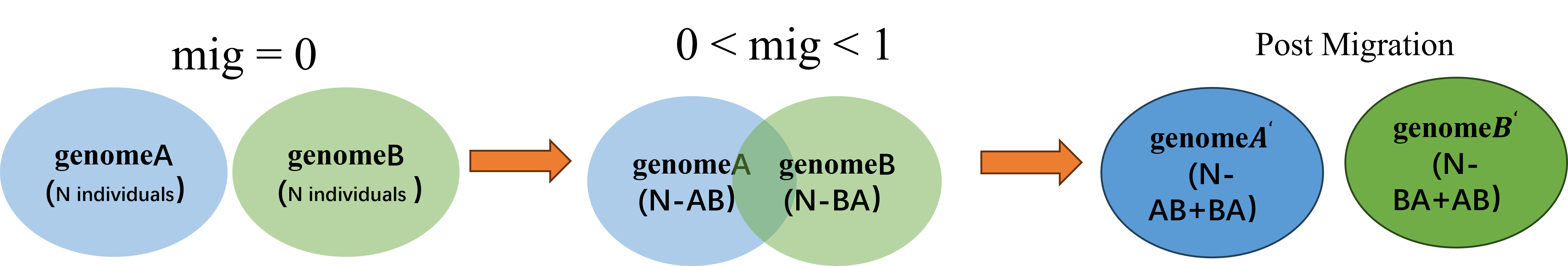}
    \caption{Schematic illustration of subpopulations A and B before and after migrations.}
    \label{fig:sample}
\end{figure}

Since migration is not implemented as a built-in evolutionary operator in FFPopSim~\cite{zanini2012ffpopsim}, we incorporated this process explicitly using scripts. Specifically, we used FFPopSim to simulate within-population evolutionary dynamics. After each generation, we then randomly selected fractions of individuals $m_{A\to B}$ and $m_{B\to A}$ from subpopulations $A$ and $B$, respectively, and exchanged them with the other subpopulation. In this way, migration was implemented through explicit exchange of individuals between subpopulations. An illustration of the migration process between the two subpopulations is shown in Fig.~\ref{fig:sample}.

We varied two primary control parameters: the migration rates and the mean additive fitness coefficient $\langle f_i\rangle$. 
The migration rate $m$ directly controls the level of individual exchange between the two subpopulations and thus represents the most relevant parameter for characterizing migration. 
The mean additive fitness coefficient $\langle f_i\rangle$ sets the overall strength of additive selection and affects whether the populations can remain in QLE-like states under the chosen recombination rate. 
Other evolutionary parameters, including the mutation rate, recombination rate, and variance of epistatic effects were fixed to values previously shown to support QLE-like behavior~\cite{zeng2020inferring, zeng2021inferring}.
Fixed parameters are listed in Table~\ref{tab:default_params}, and the scanned parameter ranges are summarized in Table~\ref{tab:varied_params}.


For each parameter set, the simulations were first run for an equilibration period, allowing the two subpopulations to reach stationary QLE-like states. Population samples were then collected after equilibration for fitness-parameter inference. The high recombination rate suppressed strong linkage disequilibrium, hence population statistics were mainly characterized by allele frequencies and weak pairwise correlations, as expected in the QLE regime. 

\section{Results}
Next, we used evolutionary simulations with known ground-truth fitness parameters, as described above, to test our inference framework.
We inferred fitness parameters separately from the sampled data of subpopulations $A$ and $B$. Additive terms $f_i$ were inferred using \eqref{eq:fi-inference-formula-v1} and epistatic couplings $f_{ij}$ were inferred using \eqref{eq:fij-inference-formula-v1}. As a comparative benchmark, we also inferred fitness parameters using traditional KNS expressions without migration. We first analyze the effects of symmetric and asymmetric migration, then we examine how the strength of selection affects inference accuracy. To quantify differences between inferred and true fitness parameters, we computed root mean squared errors (RMSE),
\begin{equation}
    \varepsilon = \sqrt{\frac{1}{B}\sum_{\alpha=1}^{B}\left(f_\alpha^*-f_\alpha\right)^2},
    \label{eq:epsilon}
\end{equation}
where $f_\alpha^*$ and $f_\alpha$ denote the inferred and true fitness parameters, respectively, and $B$ is the number of parameters. 

\subsection{Fitness inference under migration}

We begin by considering two populations with different additive selection strengths, $\langle f_i\rangle_A=0$ and $\langle f_i\rangle_B=0.001$, to test whether our approach can distinguish intrinsic fitness differences from mixing due to migration. 

\begin{figure*}[ht!]
    \centering
    \includegraphics[width=0.4\linewidth]{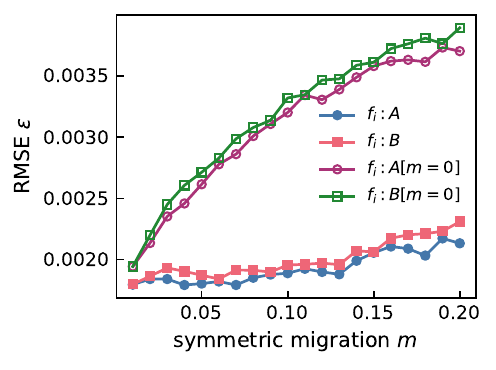}
    \put(-50,138){(a)}
    \includegraphics[width=0.4\linewidth]{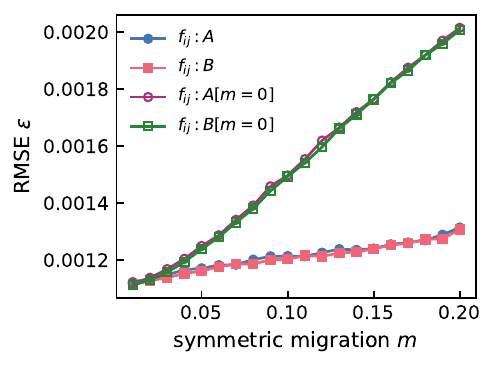}
    \put(-50,138){(b)}\\
    \includegraphics[width=0.4\linewidth]{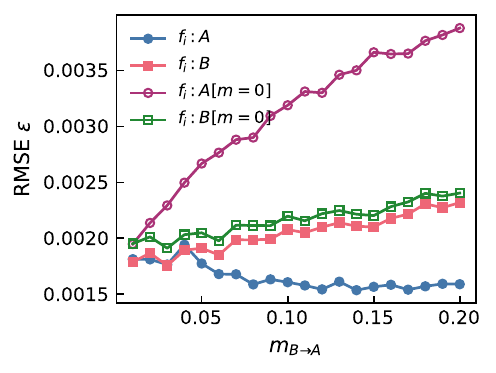}
    \put(-50,138){(c)}
    \includegraphics[width=0.4\linewidth]{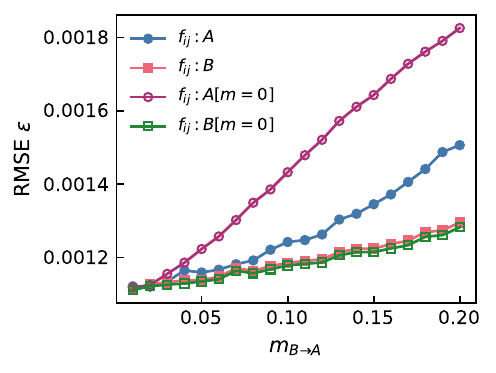}
    \put(-50,138){(d)}
    \caption{Effects of migration on the inferences of fitness parameters in two subpopulations with different selection strengths. The true fitness parameters of $A$ and $B$ are sampled from Gaussian distributions with $\langle f_i\rangle_A=0$ and $\langle f_i\rangle_B=0.001$, while the other evolutionary parameters are fixed as listed in tab.~\ref{tab:default_params}. 
    Panels (a) and (b) show the RMSE $\varepsilon$ of the inferred additive fitness terms $f_i$ and epistatic couplings $f_{ij}$ under symmetric migration, with $m_{A\to B}=m_{B\to A}=m$. 
    Panels (c) and (d) show the corresponding results under asymmetric migration, where the migration rate from population A to B is fixed at $m_{A\to B}=0.01$, while the reverse migration rate $m_{B\to A}$ is varied.  
    Solid symbols denote results obtained using the migration inference theory, whereas open symbols denote the no-migration approximation with the migration terms neglected by setting $m=0$. 
    Under symmetric migration, migration theory reduces the inference error for both additive and epistatic terms. 
    Under asymmetric migration, the improvement depends more strongly on the population and migration regime, reflecting the unequal perturbation of allele-frequency and pairwise-correlation structures by directional gene flow.
Note that by tab.~\ref{tab:default_params}
the typical size of additive fitness
($\sigma_1$) is $0.006$,
and the typical size of 
epistatic fitness ($\sigma_2$) is $0.004$.
Hence the relative errors of inference are
mostly much less than unity, see 
Fig.~\ref{fig:scatter_migration_comparison}
for representative scatter plots.
    }
\label{fig:effects_of_migration_rate}
\end{figure*}

Fig.~\ref{fig:effects_of_migration_rate}(a) and (b) show the inference errors of $f_i$ and $f_{ij}$ under symmetric migration, with $m_{A\to B}=m_{B\to A}=m$. For both additive and epistatic fitness parameters, the no-migration approximation gives considerably larger RMSEs than the migration-aware theory over the whole range of migration rates. This result shows that even balanced gene flow can bias the inferred additive selection parameters when migration is ignored, particularly when $A$ and $B$ have different selection strengths.  
Overall, the no-migration approximation exhibits a steady increase in RMSE as the symmetric migration rate $m$ increases, whereas the migration theory maintains lower errors for both populations.

Next we consider asymmetric migration, fixing $m_{A\to B}=0.01$ and allowing $m_{B\to A}$ to vary. The corresponding results are shown in Figs.~\ref{fig:effects_of_migration_rate}(c) and (d).  
We find that the accuracy of additive fitness inference strongly depends on the subpopulation. For subpopulation $A$, whose outgoing migration rate is fixed at the relatively small value $m_{A\to B}=0.01$, error rates remain low as $m_{B\to A}$ increases, so long as migration is accounted for. In contrast, the no-migration approximation produces a steadily increasing error for $A$, indicating the increasing influx from $B$ is incorrectly interpreted as selection when migration is neglected. 
For subpopulation $B$, which experiences a steady and low rate of incoming migration, eq.~\eqref{eq:fi-inference-formula-v1} provides a smaller but still discernible advantage compared to the no-migration approximation.

The inference of epistatic terms under asymmetric migration is more subtle, as shown in Fig.~\ref{fig:effects_of_migration_rate}(d). For subpopulation $A$, which experiences greater incoming migration, eq.~\eqref{eq:fij-inference-formula-v1} provides lower RMSEs than the no-migration approximation.
Errors increase along with $m_{B\to A}$, suggesting that directional gene flow progressively perturbs the pairwise correlations used to infer $f_{ij}$. 
For subpopulation $B$, the RMSE obtained from eq.~\eqref{eq:fij-inference-formula-v1} is similar or even very slightly larger than the no-migration approximation when $m_{B\to A}$ is large. 
This behavior may reflect the limitations of the present approximations when migration is very strong. This is not necessarily surprising, as the upper end of the migration rates that we tested ($m=0.2$) are no longer obviously in the regime where migration is perturbative.

\subsection{Fitness inference under different selection strengths} 

This section shows how the selection strengths affect the inference of fitness parameters. Here the asymmetric migration rates are fixed as $m_{A\to B}=0.01$ and $m_{B\to A}=0.05$ and the mean additive terms is changed. 

\begin{figure*}[ht!]
    \centering
    \includegraphics[width=0.4\linewidth]{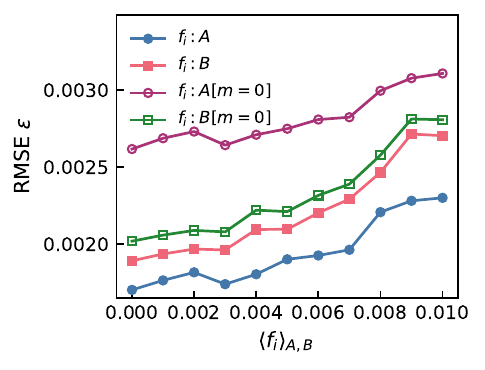}
    \put(-50,137){(a)}
    \includegraphics[width=0.41\linewidth]{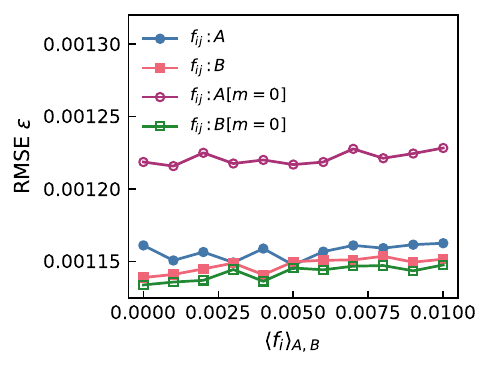}
    \put(-50,137){(b)} \\
    \includegraphics[width=0.4\linewidth]{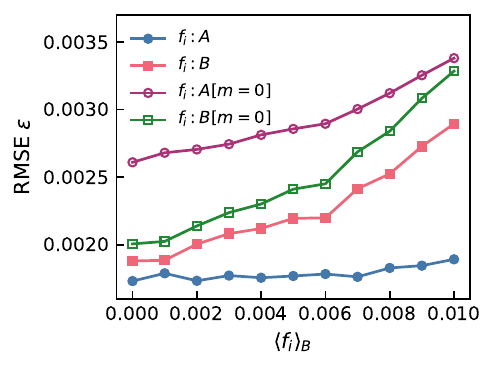}
     \put(-50,135){(c)}
    \includegraphics[width=0.41\linewidth]{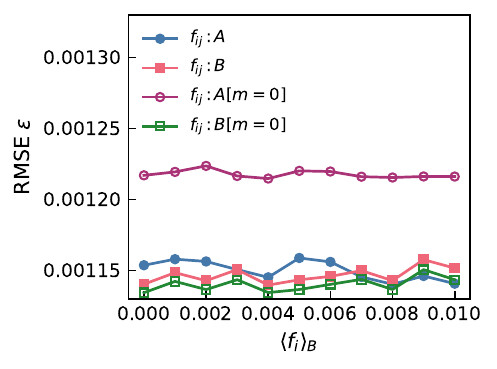}
     \put(-50,137){(d)}
    \caption{Effects of selection strength on the inference behavior of fitness parameters under asymmetric migration $m_{A\to B}=0.01$ and $m_{B\to A}=0.05$. 
    Panels (a) and (b) show the RMSE $\varepsilon$ of the inferred additive fitness terms $f_i$ and epistatic couplings $f_{ij}$, respectively, when the mean additive fitness coefficients of $A$ and $B$ are varied simultaneously, denoted by $\langle f_i\rangle_{A,B}$. 
    Panels (c) and (d) show the corresponding results when only the mean additive fitness coefficient of $B$, $\langle f_i\rangle_B$ is varied, while that of $A$ is kept fixed as $\langle f_i\rangle_A=0$. 
    The comparisons between the results from migration formulae (solids) and the no-migration version (empties) show that  the former generally yields lower RMSEs than the later approach for additive fitness inference. 
    For epistatic coupling inference, the difference between the two strategies is smaller and population dependent, eq.~\eqref{eq:fij-inference-formula-v1} gives lower errors for $A$, whereas the no-migration approximation is slightly better for $B$ with absolute small difference.
    Again from tab.~\ref{tab:default_params}
the typical size of additive fitness
($\sigma_1$) is $0.006$,
and the typical size of 
epistatic fitness ($\sigma_2$) is $0.004$.
}
\label{fig:effects_of_means_of_fis}
\end{figure*}

Fig.~\ref{fig:effects_of_means_of_fis}(a) and (b) show the inference error when $\langle f_i\rangle_A$ and $\langle f_i\rangle_B$ are varied simultaneously, denoted by $\langle f_i\rangle_{A,B}$. 
For additive fitness inference, Fig.~\ref{fig:effects_of_means_of_fis}(a) shows that the RMSE $\varepsilon$ increases with increasing $\langle f_i\rangle_{A,B}$ for both populations. This means stronger additive selection drives the populations further away from the weak-selection QLE-like regime assumed in the inference theory, makes the reconstruction of $f_i$ more difficult. 
Eq.~\eqref{eq:fi-inference-formula-v1} presents lower RMSEs than the no-migration approximation over the whole range of selection strengths, indicating the migration formula reduces the error in additive-fitness reconstruction even under asymmetric gene flow.

The corresponding results for epistatic couplings are shown in Fig.~\ref{fig:effects_of_means_of_fis}(b). Unlike the additive terms, the RMSE of the inferred $f_{ij}$ remains nearly unchanged as $\langle f_i\rangle_{A,B}$ increases. Within the present parameter range, the inference accuracy of epistatic couplings is only weakly affected by $\langle f_i\rangle_{A,B}$. 
The comparison between the two inference strategies is also more population dependent. 
For $A$, the migration theory gives slightly lower RMSEs than the no-migration approximation. 
For $B$, the no-migration approximation yields slightly smaller errors. However, the absolute difference between the two methods is very small, indicating that their performance for $f_{ij}$ inference in $B$ is comparable.

Figures~\ref{fig:effects_of_means_of_fis}(c) and (d) further show the case in which only $\langle f_i\rangle_{B}$ varies, while that of $\langle f_i\rangle_{A}=0$ and is kept fixed. 
For additive fitness inference, Fig.~\ref{fig:effects_of_means_of_fis}(c) shows a population dependent response. The RMSE in $B$ increases with $\langle f_i\rangle_B$, whereas that of $A$ remains relatively low and changes only weakly. This is expected behavior as the selection strength is varied only in $B$. In both populations, the migration inference gives lower errors than the no-migration approximation, again confirming the importance of including migration when inferring additive fitness parameters.

For epistatic inference in Fig.~\ref{fig:effects_of_means_of_fis}(d), the RMSE remains almost independent of $\langle f_i\rangle_B$. This indicates that changes in $\langle f_i \rangle_B$ have subtle effect on the reconstruction of pairwise epistatic couplings. 
Similar to Fig.~\ref{fig:effects_of_means_of_fis}(b), the migration theory presets lower errors for $A$, whereas for $B$ the no-migration approximation is slightly better or comparable.

\subsection{Representative scatter plots of inferred fitness}

To visualize the inference errors of fitness parameters, certain representative scatter plots are presented to compare the inferred fitness parameters with their true input values under migration conditions. 

\begin{figure*}[ht!]
    \centering
    \includegraphics[width=0.4\linewidth]{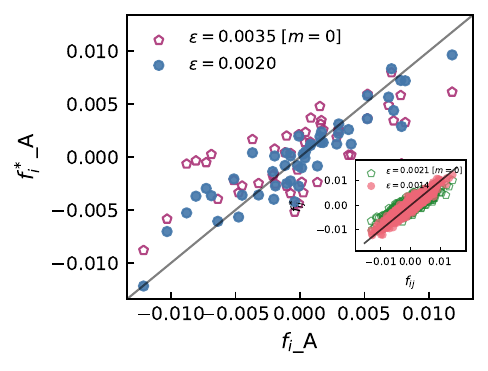}
    \put(-140,115){(a)}
    \includegraphics[width=0.4\linewidth]{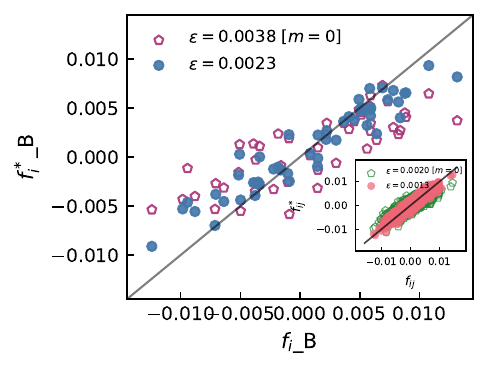}
    \put(-145,115){(b)} \\
    \includegraphics[width=0.4\linewidth]{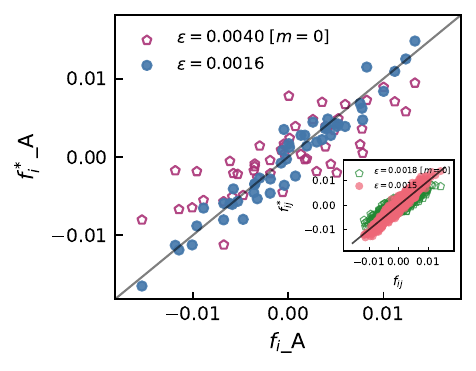}
    \put(-140,115){(c)}
    \includegraphics[width=0.4\linewidth]{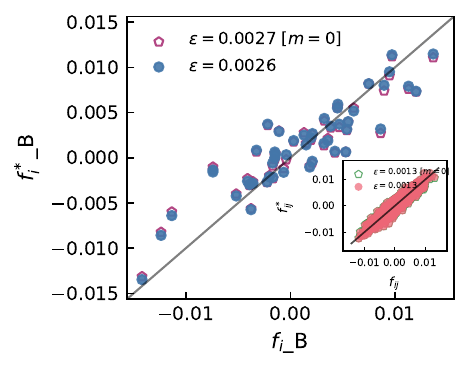}
    \put(-145,115){(d)}
    \caption{
    Representative scatter plots comparing the migration inference theory (solids) and the no-migration approximation (empties) with heterogeneous selection strengths $\langle f_i \rangle_A=0$ and $\langle f_i \rangle_B=0.001$. 
    The main panels show the inferred $f_i^*$ versus the ground-truth of $f_i$ for $A$ and $B$, while the insets for the corresponding results of $f_{ij}$. 
    Panels (a) and (b) correspond to symmetric migration with $m_{A\to B}=m_{B\to A}=0.2$, while panels (c) and (d) for asymmetric migration with $m_{A\to B}=0.01$ and $m_{B\to A}=0.2$. Under symmetric migration, migration formula yields additive fitness estimates that are closer to the diagonal than those obtained from the no-migration approximation. 
    For asymmetric migration, the reconstruction remains more population dependent, with the migration theory improving the inference in $A$, while the difference between the two strategies is smaller in $B$.
}
\label{fig:scatter_migration_comparison}
\end{figure*}

Fig.~\ref{fig:scatter_migration_comparison} shows the results for two subpopulations with heterogeneous selection strengths, $\langle f_i\rangle_A=0$ and $\langle f_i\rangle_B=0.001$. 
The main panels compare the inferred additive fitness terms $f_i^*$ with the true values $f_i$, while the insets show the corresponding comparison for the epistatic couplings. 
Filled symbols represent the migration-aware inference, and open symbols represent the no-migration approximation with $m=0$.

For symmetric migration, shown in Figs.~\ref{fig:scatter_migration_comparison}(a) and (b), the migration inference produces points that are more closely clustered around the diagonal line than the no-migration approximation. 
This improvement is also reflected in the smaller RMSE values shown in the panels. In both population $A$ and $B$, the RMSE of additive fitness inference decreases comparing with those from the no-migration approximation. 
These results provide a direct visual confirmation that balanced gene flow can still bias the inferred additive fitness terms if migration is ignored, even when two subpopulations have different selection strengths.

For asymmetric migration, shown in Figs.~\ref{fig:scatter_migration_comparison}(c) and (d), the effect becomes more population dependent. 
In $A$, the migration inference clearly reduces the scattering around the diagonal, with the RMSE decreasing from $\epsilon=0.0040$ to $\epsilon=0.0016$. 
This indicates that explicitly accounting for migration can effectively correct the inference of additive fitness in this subpopulation. 
In $B$, however, the difference between the two strategies is much smaller, with RMSE values of $\epsilon=0.0027$ for the no-migration approximation and $\epsilon=0.0026$ for the migration inference theory. 
Thus, under asymmetric migration, the benefit of including migration depends on the migration direction and on which subpopulation is considered.

The insets show that the inferred epistatic couplings $f_{ij}$ also follow the same general comparison. 
As epistatic inference relies on pairwise correlations, it is sensitive to migration-induced mixing between subpopulations. 
The scatter plots indicate that the migration inference generally captures the epistatic couplings with comparable or improved accuracy relative to the no-migration approximation, although the magnitude of improvement is smaller and more dependent on the migration pattern than for the additive terms.

\subsection{Reconstruction of inter-population fitness differences}

\begin{figure*}[ht!]
    \centering
    \includegraphics[width=0.4\linewidth]{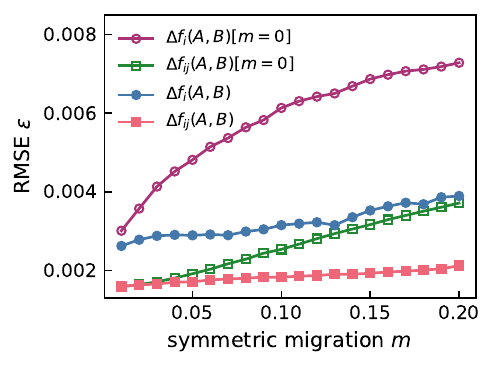}
    \put(-30,135){(a)}
    \includegraphics[width=0.4\linewidth]{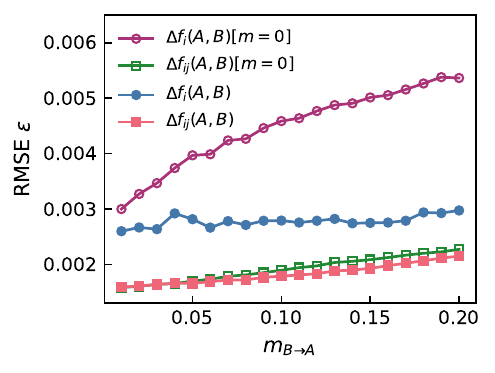}
    \put(-30,136){(b)}
    \caption{Inference accuracy for the inter-population differences in fitness parameters under symmetric and asymmetric migration. $A$ and $B$ have heterogeneous selection strengths, with $\langle f_i\rangle_A=0$ and $\langle f_i\rangle_B=0.001$. Here the RMSE $\varepsilon$ is calculated between the true differences of fitness parameters in populations $A$ and $B$ and the
differences of the corresponding inferred fitness parameters using      
\eqref{eq:fi-inference-formula-v1}
and
\eqref{eq:fij-inference-formula-v1}.
    Panel (a) shows the results under symmetric migration, with $m_{A\to B}=m_{B\to A}=m$. 
    Panel (b) shows the results under asymmetric migration, where $m_{A\to B}=0.01$ is fixed and $m_{B\to A}$ is varied. Here the blue circles and red squares denote the RMSEs of the inferred inter-population differences in additive fitness terms $\Delta f_i(A,B)$, and epistatic couplings $\Delta f_{ij}(A,B)$, obtained using the migration inference theory. Purple circles and green squares denote the corresponding results obtained from the no-migration approximation with $m=0$. The migration theory yields lower RMSEs for the reconstruction of inter-population differences, especially for additive fitness terms, whereas the no-migration approximation leads to larger errors as the migration rate increases.
From tab.~\ref{tab:default_params}
the typical size of additive fitness
($\sigma_1$) is $0.006$,
and the typical size of 
epistatic fitness ($\sigma_2$) is $0.004$,
hence the displayed inferred differences are typically here much smaller.
 }
\label{fig:inter_population_difference}
\end{figure*}

Beyond inferring the fitness parameters within each subpopulation, it is also important to determine whether the inferred parameters can preserve the differences between the fitness landscapes of $A$ and $B$. 
Therefore this section examine the inter-population differences in additive fitness terms and epistatic couplings, denoted as $\Delta f_i(A,B)$ and $\Delta f_{ij}(A,B)$, and compare their inferred values with the corresponding true differences.  
Figure~\ref{fig:inter_population_difference} compares the true inter-population differences, $\Delta f_i(A,B)$ and $\Delta f_{ij}(A,B)$, with the corresponding inferred differences under symmetric and asymmetric migration. 

Under symmetric migration, shown in Fig.~\ref{fig:inter_population_difference}(a), the migration theory gives clearly lower RMSEs than the no-migration approximation, especially for the additive fitness difference $\Delta f_i(A,B)$. 
The error of the no-migration approximation increases rapidly with the migration rate, indicating that ignoring migration can distort the inferred difference between the two additive fitness landscapes. 
For the epistatic difference $\Delta f_{ij}(A,B)$, the improvement is more moderate but remains visible.

Under asymmetric migration, shown in Fig.~\ref{fig:inter_population_difference}(b), the migration theory also yields lower errors for the inferred additive difference, while the no-migration approximation again shows increasing error as $m_{B\to A}$ becomes larger. 
For the epistatic difference, the two strategies provide closer RMSE values, suggesting that the reconstruction of $\Delta f_{ij}(A,B)$ is less strongly affected by migration than that of $\Delta f_i(A,B)$ in this parameter regime.

\FloatBarrier

\section{Conclusion}
We have for the first time considered Quasi-Linkage Equilibrium in the presence of migration.
We have developed an extension of the Kimura-Neher-Shraiman theory which allows to infer additive and (pairwise) epistatic fitness parameters from the distribution of genotypes in a population, and we have checked these inference formulae in simulations.

We have set up a fairly simple model to take in account migration in a QLE phase. However, the model does not only take in account the case of a symmetric migration between two populations $A$ and $B$, that is to say the migration rate is the same in the two direction $A\to B$ and $B\to A$, but also asymmetric migration. 
In addition to migration between the two ``islands'' or subpopulations, we have also accounted for selection, mutation and recombination within each island. As the selection depends on the fitness landscape in the surroundings of each individuals, the fitness parameters $(f_{i}^A,f_{ij}^B)$ are specific to the subpopulations on respectively island $A$ and island $B$. 
To focus on the new effect of migration we have limited ourselves to the cases where the recombination rate $r$ and the mutation rate $\mu$ are the same in the two populations. 
In our model, the size of a given population may fluctuate over time but the total number of individuals within the two subpopulations is considered fixed. This can model the struggle for survival in a constrained environment with limited resources, for instance two nearby islands in the middle of the sea.

In the QLE approach, that is to say by considering the case where the genomic distribution is similar to the distribution of a set of spins in the case of an Ising model, we found that the migration process acts as a corrective term for the dynamics of both the first and second order cumulants. First order cumulants (means) are widely used to characterize contributions of single nucleotide polymophisms to heritability and variability of complex traits, for a recent study on human data 
(347,630, 40 million loci), see 
\cite{wainschtein2025estimation}.
The results reported here strongly suggest that also epistatic fitness terms can be reliably inferred, also in the presence of migration.

We obtained a new inference formula for epistatic fitness~\eqref{eq:Inferred_epistasis_pop_A}. This formula, together with the equivalent formula for island $B$, suggest that it is possible for both islands to be in a QLE phase at the same time, without those phases being the same. In that way, the two phases are seen in mutual influence from one another, without their states being fundamentally disrupted by this influence.

We found that migration plays an important role in the inference of fitness parameters in structured populations.  The migration theory reduces reconstruction errors for both additive fitness terms $f_i$ and epistatic couplings $f_{ij}$ under symmetric migration, while under asymmetric migration the improvement becomes more dependent on the migration direction, the affected subpopulation, and the type of inferred parameter. 
Selection strength mainly affects additive fitness inference, with the RMSE of $f_i$ increasing as $\langle f_i\rangle$ becomes larger, whereas the inference of $f_{ij}$ remains relatively insensitive to $\langle f_i\rangle$ in the parameter range considered. 
Moreover, accounting for migration helps recover not only population specific fitness parameters, but also the differences in fitness landscapes between populations.

In this study, we advance research further by for the first time incorporating bidirectional migration as the fifth evolutionary force into the defining master equation and cumulant dynamics. We derive new inference formulas that enable simultaneous estimation of additive and epistatic fitness in two subpopulations. By extending FFPopSim and coupling it with the all-time nMF algorithm, we evaluate the recovery accuracy across the parameter space ($\mu$, $r$, $m$, $\sigma_1$, $\sigma_2$). Within this space, 
moderate mutation and recombination rates provide sufficient genetic variability while maintaining significant linkage signals; low migration rates prevent gene flow from excessively homogenizing allele frequencies; moderate additive effect strength ($\sigma_1$) ensures that selection signals remain above noise levels; and low epistatic strength ($\sigma_2 \approx 0.004$) prevents higher-order interactions from overwhelming detectable pairwise associations.

In short, we have extended an important theory of population genetics from the idealized setting of 
a single, randomly mating population
to a more realistic scenario of a population divided into subpopulations and migration of individuals between these different habitats.
We show that the QLE state is robust against these extensions, and that different QLE states with different environments can co-exist although individuals move between them. We show that a the \textit{difference} of fitness in two different environments can be reliably inferred using a precise expression for this difference. 

\section*{Acknowledgement}
The work of H.L.Z. was sponsored by the National Natural Science Foundation of China (Grant No. 11705097), the China Scholarship Council (Grant No. 202508320441), and the Natural Science Foundation of Nanjing University of Posts and Telecommunications (Grants No. 221101 and No. 222134). 

\section{Appendices}
\subsection{Parameter settings} \label{app:paramater_setting}
Tables~\ref{tab:default_params} and \ref{tab:varied_params} summarize the simulation parameters used in FFPopSim. 
The default settings, including the number of loci, carrying capacity, number of generations, recombination model, and crossover rate, are listed in Table~\ref{tab:default_params} and were kept fixed unless otherwise specified. 
In particular, a relatively high crossover rate was used to maintain the populations in a recombination-dominated, QLE-like regime.
The parameters varied in the simulations are given in Table~\ref{tab:varied_params}, including the migration rate, out-crossing rate, mutation rate, and the strengths of additive and epistatic fitness effects. 
By scanning these parameters, we systematically evaluated how migration and other evolutionary factors affect the accuracy of additive and epistatic fitness-parameter inference.
\begin{table}[ht!]
\centering 
\caption{Main default parameters of FFPopSim used in simulations} 
\label{tab:default_params}
\begin{tabular}{|l|l|}
\hline 
\textbf{Parameter} & \textbf{Value} \\
\hline 
number of loci ($L$) & 50 \\
\hline
carrying capacity ($N$) & 1000 \\
\hline
generation & 1000 \\
\hline
recombination model & CROSSOVERS \\
\hline
crossover rate ($\rho$) & 1.0 \\
\hline
mutation rate ($\mu$) & 0.01 \\
\hline
out-crossing rate ($r$) & 0.5\\
\hline 
standard deviation of additive fitness  $\sigma_1$ & $0.006$ \\
\hline
standard deviation of epistatic fitness $\sigma_2$ &  $0.004$ \\
\hline
mean of epistatic fitness &  $0$ \\
\hline
\end{tabular}
\end{table}

\begin{table}[ht!]
\centering
\caption{Varied parameters of FFPopSim used in simulations} 
\label{tab:varied_params}
\begin{tabular}{|l|l|} 
\hline
\textbf{Parameter} & \textbf{Value} \\
\hline
 migration rate ($m$) & $[0.01, 0.2]$ \\
\hline
mean of additive fitness &  $[0, 0.01]$ \\
\hline
initial genotypes & binary random numbers \\
\hline
\end{tabular}
\end{table}

\subsection{Quasi-Linkage-Equilibrium } \label{sec:Quasi-Link-Equilibrium}
This section contains more details on Quasi-Linkage Equilibrium (QLE) to supplement the discussion in the main body of the paper. The notation and content partly overlaps with the presentation in \cite{dichio2023statistical}.
\subsubsection{Definition of QLE and Physical Analogy}
The concept of Quasi-Linkage Equilibrium state was originally introduced by Kimura in 1965 for two-locus two-allele models with selection, mutation and recombination in evolving populations of finite size~\cite{kimura1965attainment}. It was  developed further by Neher and Shraiman for multi-locus two-allele models~\cite{neher2009competition} and extended to multi-locus multi-allele models by Gao \textit{et al.} \cite{gao2019dca}.
We will describe KNS by adopting the following simplification: by “genotype”, we will always mean one genome out of all possible genomes of the same length. Additionally, we make the following simplifying assumptions:
\begin{enumerate}
    \item \textbf{Genomic structure.} A haploid genome is a vector $g=(s_1, \ldots, s_L) $ of $L$ loci  $s_i$, where $i = 1, \ldots, L$. The number $L$ of loci is fixed and equal for all individual genomes. A population is a collection $\{g^A\}_{A \in A}$, where $A$ is a set of indices. Each genome $g$ appears in the population with probability $P(g)$.

    \item \textbf{Ising loci.} Loci are bi-allelic,  \textit{i.e.}, there are two alleles at each locus. They can then be coded by spin-like variables $ s_i = \pm 1 \ \forall i $.

    \item \textbf{Constant population.} When considering an isolated population, the average number of individuals in this population is fixed. This hypothesis can model, \textit{e.g.}, the struggle for survival in an environment with limited resources.

    \item \textbf{One-genome evolution.} The distribution of one genome in a population is given by a genome distribution $ P(g,t) $. This distribution evolves in time driven by three operators representing natural selection, mutations and recombination.
    That recombination can be understood in terms of only one-genome distributions is a simplifying assumption analogous to Boltzmann's \textit{Sto\ss zahlansatz}.

    \item \textbf{Coupled one-genome evolutions.} To accommodate migration between spatially separated populations, we introduce in the same paradigm one distribution over genomes for each location. Each of these distributions will fulfill the conditions above, and will the evolve also by exchanging individuals, 
\end{enumerate}

\subsubsection{Natural forces of population evolution}

The action of the three evolutionary forces is then encoded in a master equation, \textit{i.e.}, a phenomenological first-order differential equation:

\begin{equation}
\frac{d}{dt}P(g,t) = 
\left.\frac{d}{dt}P(g,t)\right|_{\text{fit}} + 
\left.\frac{d}{dt}P(g,t)\right|_{\text{mut}} + 
\left.\frac{d}{dt}P(g,t)\right|_{\text{rec}}
+ 
\left.\frac{d}{dt}P(g,t)\right|_{\text{mig}}
\label{eq:original_master_equation}
\end{equation}
In this section we discuss the first of these three terms of this master equation in turn,
and then discuss the last (new) term in the next section.

\textbf{Selection.} The model for natural selection is based on a fitness function $F$ that can be understood as proportional to the expected number of offspring of an individual of genotype $g$.
In other words, $F(g)$ expresses the propensity of a genotype to transfer its genomic material to the next generations. The explicit form of $F(g)$ defines the fitness landscape of the population.
Bypassing the complex conceptual issues around biological fitness \cite{Orr2009},
we hence suppose that the fitness of an individual depends only on its genome. The first term of eq~\eqref{eq:original_master_equation} can then be written as:
\begin{equation}
\frac{d}{dt}\Big|_{\text{fit}} P(g, t) = \big[ F(g) - \langle F\rangle_{t} \big] P(g, t).
\label{eq:selection}
\end{equation}
where $\displaystyle \langle F\rangle_{t} = \sum_{g} F(g)P(g,t)$
is the population-average fitness that ensures the normalization of $P(g,t)$. Fit individuals ($F(g)-\langle F\rangle_{t}>0$) will grow in proportion, and unfit ones ($F(g)-\langle F\rangle_{t}<0$) will decrease. Therefore, even in this simplified model, whether an individual is fit or not depends on which other individuals are present in the population. We will here only consider fitness functions with linear and pairwise interactions.
\begin{equation}
F(g) = \sum_{i} f_{i} s_{i} + \sum_{i<j} f_{ij} s_{i} s_{j},
\label{eq:fitness}
\end{equation}
where the linear coefficients $f_i$ are referred to as \textit{additive} contributions to fitness and the quadratic (pairwise) coefficients $f_{ij}$ are \textit{epistatic} contributions to fitness. 
A constant fitness term as stated in the main body of the paper would 
cancel in
\eqref{eq:selection}.
The indicator variables $s_i$ being dimension-less, the dimensions of 
$f_i$ and $f_{ij}$ are $[t^{-1}]$.
From the mathematical point of view, other 
possibilities to represent fitness have been explored in the literature, see \cite{peliti1997introduction}
and references therein, but have not been considered here.

\textbf{Mutation.} The model for mutations is single-locus swaps $ s_i \to -s_i $. We hence do not take into account mutations that consist in insertions or deletion of the genomic chain. In mathematical terms an operator $ M_{i} $ acts on a genomic sequence by swapping the $i$-th bi-allelic gene, i.e. $ M_{i}(g)=(-s_i,s_{\backslash i}) $. Let $\mu$ be the tunable mutation rate, constant in time and the same for all loci, it has dimensions $[t^{-1}]$. The mutation term in the master equation then takes the simple form
\begin{equation}
\left. \frac{d}{dt} \right|_{\text{mut}} P(g,t)=\mu \sum_{i=1}^L\left[P(M_{i}g,t)-P(g,t)\right].
\label{eq:mutation}
\end{equation}

\textbf{Recombination.} The model for recombination is that two parents $g^{(1)},g^{(2)}$ mix their genomic sequences and give birth to two new individuals $g, g^{\prime}$.
It is convenient to introduce a set of random variables $\{\xi_{i}\}$ to describe recombination by defining a crossing-over pattern \cite{NeherShraiman2011,gao2019dca,dichio2023statistical}. Consider the allele at locus $i$ of the new individual $g$, if it has been inherited from $g^{(1)}$, then $\xi_{i}=1$, while if it comes from $g^{(2)}$, then $\xi_{i}=0$. The sequence $g^{\prime}$ is simply complementary to $g$. In symbols, $g,g^{\prime}$ can be written as
\begin{equation}
\begin{split}
g: \; s_{i} &= \xi_{i} s_{i}^{(1)} + (1 - \xi_{i}) s_{i}^{(2)}, \\
g^{\prime}: \; s_{i}^{\prime} &= (1 - \xi_{i}) s_{i}^{(1)} + \xi_{i} s_{i}^{(2)}.
\label{eq:new individuals}
\end{split}
\end{equation}

Each different crossover pattern $\{\xi_{i}\}$ comes with a probability $C(\bm{\xi})$. Let $r$ be the tunable overall recombination parameter, dimensions $[t^{-1}]$. Under the simplifying assumption that any genome pair has the same recombination rate $r$, population where any individual is equally likely to interact with anyone else, the recombination term can in general be written as a kind of collision term
\begin{equation}
\left. \frac{d}{dt} \right|_{\text{rec}} P(g,t) = r \sum_{\xi, g'} C(\xi) \left[ P_2(g^{(1)}, g^{(2)}, t) - P_2(g, g', t) \right].
\label{eq:recombination}
\end{equation}
where $g^{(1)}, g^{(2)}$ are found by inverting eq~\eqref{eq:new individuals}
and where $P_2$ is the two-genome distribution. In this scheme the evolution of the two-genome distribution would involve the three-genome distribution, and so on; a truncation at some level is necessary.
The sum \eqref{eq:recombination} runs over all possible recombination patterns and all possible sequences $g^{\prime}$, 

To close the equations we assume that the two-genome distributions factorize:
$$
P_2(g_A, g_B) = P(g_A) P(g_B) .
$$
which by above we have taken to be part 
of the definition of the QLE phase. 
Inserting this assumption in eq~\eqref{eq:recombination}, we get
\begin{equation}
\left. \frac{d}{dt} \right|_{\text{rec}} P(g,t) = r \sum_{\xi, g^{\prime}} C(\xi) \left[ P(g^{(1)}, t) P(g^{(2)}, t) - P(g, t) P(g^{\prime}, t) \right]. 
\label{eq:recombination_1}
\end{equation}
The final expression for the master equation
describing evolution due to selection, mutations and recombination 
can now be obtained by inserting expressions~\eqref{eq:selection}, \eqref{eq:mutation} and \eqref{eq:recombination_1} into \eqref{eq:original_master_equation}:
\begin{equation}
\begin{split}
\frac{d}{dt}P(g,t) &= \left[F(g) - \langle F \rangle_t \right] P(g,t) \\
&\quad + \mu \sum_{i=1}^{L} \left[ P(M_i g,t) - P(g,t) \right] \\
&\quad + r \sum_{\xi, g^{\prime}} C(\xi) \left[ P(g^{(1)},t) P(g^{(2)},t) - P(g, t) P(g^{\prime},t)\right]. 
\label{eq:master_equation}
\end{split}
\end{equation}

\subsection{Migration term in the master equation}
We here continue on the fourth term in \eqref{eq:original_master_equation}.
At $t+dt$, the number of individuals with genome g in the population $A$ will be: 
\begin{align*}
n_{A}(g,t+dt)&=n_{A}(g,t)+\left(m_{B\to A}n_{B}(g,t)-m_{A\to B}n_{A}(g,t)\right)dt\\[3pt]
&=n_{A}(g,t)+\left(m_{B\to A}N_{B}(t)P_{B}(g, t)-m_{A\to B}N_{A}(t)P_{A}(g,t)\right)dt.
\end{align*}
At the same time, the size of population $A$ evolves as: 
\begin{equation}
    \label{eq:size_alpha}
    N_{A}(t+dt)=N_{A}(t)+\left(m_{B\to A}N_{B}(t)-m_{A\to B}N_{A}(t)\right)dt.
\end{equation}
We can write (this expression formally mixes orders in $dt$ but is stated for clarity) 
\begin{equation*}
    P_A(g,t+dt)=\frac{n_A(g,t)}{N_A(t)+\left(m_{B\to A}N_B(t)-m_{A\to B}N_A(t)\right)dt}+\frac{m_{B\to A}N_B(t) P_B(g,t)-m_{A\to B}N_A(t) P_A(g,t)}{N_A(t)+\left(m_{B\to A}N_B(t)-m_{A\to B}N_A(t)\right)dt}dt
\end{equation*}
which to first order in $dt$ gives
\begin{equation*}
    \begin{aligned}
     P_A(g,t+dt)=&
     &\frac{n_A(g,t)}{N_A(t)} + dt\left[m_{B\to A} \frac{N_B(t)}{N_A(t)}\left(P_B(g,t)-P_A(g,t)\right) + m_{A\to B}\left(P_A(g,t) -P_A(g,t)\right)\right]
     \end{aligned}
\end{equation*}
Inserting $P_A(g)=n_A(g)/N_A$ we have
\begin{equation}
    P_A(g,t+dt)=P_A(g,t)+m_{B\to A}\frac{N_B}{N_A}\left[P_B(g,t)-P_A(g,t)\right]dt.
    \label{eq:evolution_proba}
\end{equation}
Which yields
\begin{equation}
    \label{eq:derivative_mig}
    \left. \frac{d}{dt} \right|_{\text{mig}}P_{A}(g) =m_{B\to A}\frac{N_{B}}{N_{A}}\left[P_{B}(g)-P_{A}(g)\right].
\end{equation}
Perhaps surprisingly, 
the evolution of $P_A$ by \eqref{eq:derivative_mig} does not depend on
the migration rate from $A$ to $B$,
that is
$m_{A\to B}$, but only on the migration rate from $B$ to $A$. That is because the migration rate from $A$ to $B$ impacts both the size of population $A$ and the number of individuals carrying the genome $\mathbf{g}$ in $A$ proportionally, leaving $P_A(g)$ unchanged.

\subsection{Inference of epistatic fitness in presence of migration} \label{sec:Inference}
In this section we fill in details on the derivation of the inference formulae for epistatic fitness in two populations $A$ and $B$ which exchange individuals (migration) and where the fitness functions are different. We focus on epistatic fitness as this leads particularly compact formulae.

We seek here to get an inference formula, which is a formula linking the fitness parameters ($f_{i}$, $f_{ij}$) with the Ising parameters ($h_{i}$, $J_{ij}$) constituent of the definition of the QLE phase as defined in eq.~\eqref{eq:QLE}. 
To link fitness parameters to the Ising parameters, we consider the time derivative of the logarithm of the probability. We here only consider the population A, equations for population $B$ are obtained by exchanging the indices $A \leftrightarrow B$. 
The starting point is thus
\begin{equation}
\frac{d}{dt}\ln P_{A}(g)=-\frac{\dot{\mathcal{Z}}_{A}}{\mathcal{Z}_{A}}+\sum_{i=1}^{L}\dot{h}_{i}^{A}s_i+\sum_{i<j}\dot{J}_{ij}^{A}s_{i}s_{j}. 
\label{eq:partition_function_log_deriv_app} 
\end{equation}
where the time derivatives of the parameters 
$h_{i}$ and $J_{ij}$ appear. We emphasize again that from the perspective of this paper these parameters are effective descriptions of a distribution of genotypes which depend on underlying parameters as we are in the process of determining. These underlying parameters may depend explicitly on time or the relation can lead to auxiliary dependence on a time-dependent state, and in both cases $h_{i}$ and $J_{ij}$ can depend on time. In particular, $h_{i}$ will typically depend on time.  
By definition
\begin{equation}
\frac{d}{dt}\ln P_{A}(g)=\frac{1}{P_{A}(g)}\frac{dP_{A}(g)}{dt}.
\label{eq:derivative_probability}
\end{equation}
and using \eqref{eq:partition_function_log_deriv_app}
on the left hand side and 
\eqref{eq:master_equation}
and
\eqref{eq:derivative_mig}
on the right hand side
we have an equality between, on one hand, the time derivative of the field parameters ($\dot{h}_{i}$, $\dot{J}_{ij}$), and on the second hand, the field parameters ($h_{i}$, $J_{ij}$) and the fitness parameters ($f_{i}$, $f_{ij}$). More explicitly 
we have the master equation
\begin{align}
\frac{dP_A(g)}{dt}\frac{1}{P_A(g)}
&=
[F_A(g)-\langle F_A\rangle_t]
\quad \text{(selection)} \nonumber \\
&\quad
+
\mu\sum_{i=1}^L
\left[
\frac{P_A(M_i g)}{P_A(g)} - 1
\right]
\quad \text{(mutation)} \label{eq:derivate_master_eq} \\
&\quad
+
r\sum_{\xi,g'}
C(\xi)P_A(g')
\left[
\frac{
P_A(g^{(1)})P_A(g^{(2)})
}{
P_A(g)P_A(g')
}
-1
\right]
\quad \text{(recombination)} \nonumber \\
&\quad
+
m_{B\to A}\frac{N_B}{N_A}
\left[
\frac{P_B(g)}{P_A(g)} -1
\right]
\quad \text{(migration)}. \nonumber
\end{align}
The first term is equal to
\begin{equation}
\text{(selection)} = \sum_{i=1}^L f_i^A s_i + \sum_{i<j} f_{ij}^A s_i s_j. 
\label{eq:the_first_term}
\end{equation}
For the second one, we notice that in $P_A(M_i g)$, only the terms $i$ that have been flipped are different from those present in $P_A(g)$. Hence
\begin{equation}
\frac{P_A(M_i g)}{P_A(g)} = \exp\left\{ -2\left( h_i^A s_i + \sum_{\substack{j=1 \\ j \neq i}}^L J_{ij}^A s_i s_j \right) \right\}.
\label{eq:P_A(M_ig)_over_P_A(g)}
\end{equation}
We can consider that the term in the exponential is small, as we only shifted one spin between $P_{A}(g)$ and $P_{A}(M_i g)$. This enables us to approximate the exponential at first order. The second term then becomes
\begin{equation}
\text{(mutation)} = -2\mu\left[\sum_{i=1}^{L} h_i^As_i + 2\sum_{i<j} J_{ij}^A s_i s_j\right]. 
\label{eq:the_second_term}
\end{equation}

As for the third one, it has been computed in \cite{NeherShraiman2011}, see also \cite{dichio2023statistical}, by using the definition of QLE eq~\eqref{eq:QLE} and is equal to
\begin{equation}
\text{(recombination)} = r\sum_{i<j} c_{ij} J_{ij}^A \left[s_i \chi_j^A + \chi_i^A s_j - s_i s_j + \langle s_i s_j \rangle_A\right].
\label{eq:the_third_term}
\end{equation}
where $c_{ij}$ is the result of summing over $\xi$, a quantity which depends on loci $i$ and $j$ and which can be interpreted as the probability that the alleles at $i$ and $j$ are inherited from different parents. Hence $c_{ij}$ should be close to zero for pairs of loci that are close because unless a recombination event happens between the loci the alleles will be inherited from the same parent. On the other hand, for overall recombination rate $r$ high enough and pairs of loci far enough apart, $c_{ij}$ should be about one half; the alleles could be either inherited from the same parent of from different parents, with about equal probability.

Lastly, we need to compute the fourth term. For simplicity, we write

\begin{equation}
\sum_{i=1}^{L} h_i^As_i + \sum_{i<j} J_{ij}^A s_i s_j = E_A, \quad \text{and} \quad \sum_{i=1}^{L} h_i^Bs_i + \sum_{i<j} J_{ij}^B s_i s_j = E_B.
\label{eq:simplicity}
\end{equation}
Then
\begin{equation}
\frac{P_B(g)}{P_A(g)} - 1 = \frac{\mathcal{Z}_A}{\mathcal{Z}_B} e^{E_B-E_A} - 1.
\label{eq:Formula_transformation}
\end{equation}
Assuming $\sqrt{L} \cdot \sigma(\{\Delta h_i\}) \ll 1 \quad \text{and} \quad L \cdot \sigma(\{\Delta J_{ij}\}) \ll 1$,  the difference $E_B-E_A$ can be considered to be small enough so we can approximate it at first order:
\begin{equation}
\text{(migration)} = m_{B\to A}\frac{N_B}{N_A} \left[\frac{\mathcal{Z}_A}{\mathcal{Z}_B}(E_B-E_A) + \frac{\mathcal{Z}_A}{\mathcal{Z}_B} - 1\right]. 
\label{eq:the_last_term}
\end{equation}
Now, by setting \eqref{eq:partition_function_log_deriv_app} equal to \eqref{eq:derivate_master_eq}, we get
\begin{align}
-\frac{\dot{\mathcal{Z}}_A}{\mathcal{Z}_A}
+
\sum_i \dot{h}_i^A s_i
+
\sum_{i<j} \dot{J}_{ij}^A s_i s_j
&=
\sum_i f_i^A s_i
+
\sum_{i<j} f_{ij}^A s_i s_j
\quad \text{(selection)} \nonumber
\\
&-
2\mu
\left(
\sum_i h_i^A s_i
+
2\sum_{i<j} J_{ij}^A s_i s_j
\right)
\quad \text{(mutation)}
\label{eq:Formula_transformation_1}
\\
&+
r
\sum_{i<j}
c_{ij} J_{ij}^A
\left(
s_i \chi_j^A
+
\chi_i^A s_j
-
s_i s_j
+
\langle s_i s_j \rangle_A
\right)
\quad \text{(recombination)} \nonumber
\\
&+
m_{B\to A}\frac{N_B}{N_A}
\Bigg[
\frac{\mathcal{Z}_A}{\mathcal{Z}_B}
\left(
\sum_i (h_i^B-h_i^A)s_i
+
\sum_{i<j}(J_{ij}^B-J_{ij}^A)s_is_j
\right)
+
\frac{\mathcal{Z}_A}{\mathcal{Z}_B}
-1
\Bigg].
\quad \text{(migration)}
\nonumber
\end{align}
By collecting together terms with the same monomials in $s_i$ and $s_is_j$, one gets the two equations where one involves the additive fitness parameter $ f_i^A $ and one involves epistatic fitness parameter $f_{ij}^A$:
\begin{equation}
\dot{h}_i^A 
= {f}_i^{A,*}
- 2\mu h_i^A + r\sum_{j}c_{ij} J_{ij}^A \chi_j^A  - m_{B\to A} \frac{N_B}{N_A} \frac{\mathcal{Z}_A}{\mathcal{Z}_B} (h_i^A - h_i^B)
\label{eq:additive_fitness_of_popA}
\end{equation}
and
\begin{equation}
\dot{J}_{ij}^A =
f_{ij}^{A\note{,*}} - 4\mu J_{ij}^A- r c_{ij} J_{ij}^A  - m_{B\to A}\frac{N_B}{N_A} \frac{\mathcal{Z}_A}{\mathcal{Z}_B} \left( J_{ij}^A - J_{ij}^B \right).
\label{eq:Jij_after_migration}
\end{equation}

When recombination and mutation dominate over migration the second equation is a relaxation equation for $J_{ij}^A$. Thus, equation \eqref{eq:Jij_after_migration} together with its counterpart for population B yields the matrix equation, where the nonlinearity is hidden in the dependence of 
$(\mathcal{Z}_A,\mathcal{Z}_B)$
on $(J_{ij}^A,J_{ij}^B)$:
\begin{equation}
    \dot{\mathbf{J}}=\mathbf{f}-\mathcal{M}\mathbf{J}
\label{eq:matrix_epistatis}
\end{equation}
\begin{equation*}
    \mathbf{J}=\begin{pmatrix}
        \displaystyle J_{ij}^A \\[3pt]
        \displaystyle J_{ij}^B
    \end{pmatrix}, \qquad 
    \mathbf{f}=\begin{pmatrix}
        \displaystyle f_{ij}^A \\[3pt]
        \displaystyle f_{ij}^B
    \end{pmatrix}, \qquad 
    \mathcal{M}=\begin{pmatrix}
        \Delta+\nu_{B\to A} & -\nu_{B\to A}\\[3pt]
        -\nu_{A\to B} & \Delta+\nu_{A\to B}
    \end{pmatrix}
\end{equation*}
where $\Delta= 4\mu +r c_{ij}$ and $\nu_{A\to B}=m_{A\to B}\dfrac{N_A}{N_B}\dfrac{\mathcal{Z}_B}{\mathcal{Z}_A}$.
If relaxation is sufficiently fast $\mathbf{J}$ will tend to a conditional equilibrium given by setting the left hand side of  
\eqref{eq:matrix_epistatis} to zero: 
\begin{equation}
    \mathbf{J}=\mathcal{M}^{-1} \mathbf{f}.
\label{eq:matrix_epistatis_eq}
\end{equation}
The two relaxation rates are given by the eigenvalues of $\mathcal{M}$:
\begin{equation*}
    \lambda_-=\Delta+\nu_{A\to B}+\nu_{B\to A} \qquad
    \lambda_+=\Delta,
\end{equation*}
associated to the two decay modes:
\begin{equation*}
    \mathbf{e}_-=\begin{pmatrix}
        \displaystyle \nu_{B\to A} \\[3pt]
        \displaystyle -\nu_{A\to B}
    \end{pmatrix} \qquad 
    \mathbf{e}_+=\begin{pmatrix}
        \displaystyle 1 \\[3pt]
        \displaystyle 1
    \end{pmatrix}
\end{equation*}
When the two subpopulations vary in the same direction, the common mode $e_+$ captures the shared component of the population dynamics — reflecting, for instance, common selection pressures or global demographic trends — and is unaffected by migration, since exchanging individuals between two identical population states leaves their difference unchanged. The differential mode $e_-$, by contrast, encodes the divergence between subpopulations and is directly governed by the total migration rate $\nu_{A\to B}+\nu_{B\to A}$.
Thus, migration primarily acts on the differential mode by suppressing or reshaping population divergence, while leaving the common mode unchanged. 
Physically, these two modes are analogous to center-of-mass and relative modes in a coupled two-component system. These two modes are coexisted  no matter the system is under symmetric or asymmetric migrations.
When \eqref{eq:matrix_epistatis_eq} is satisfied $(J_{ij}^A,J_{ij}^B)$ 
are together in conditional equilibrium with respect to $(f_{ij}^A,f_{ij}^B)$, the parameters $r,\mu,m_{A\to B},m_{B\to A}$
and the values $(\mathcal{Z}_A,\mathcal{Z}_B)$
and $(N_A,N_B)$. We will soon show how the ratio $\mathcal{Z}_B/\mathcal{Z}_A$ can be expressed in terms of the other parameters in a mean-field ansatz. Assuming this to be the case, we can turn the problem around, and take 
$(f_{ij}^A,f_{ij}^B)$ to be the unknowns,
$(J_{ij}^A,J_{ij}^B)$ and $(N_A,N_B)$ to be the data, and $r,\mu,m_{A\to B},m_{B\to A}$ to be assumed known by other means. This then leads to the first inference formula for epistatic fitness in the presence of migration:
\begin{equation}
f_{ij}^{A\note{,*}} =  (4\mu +r c_{ij}) J_{ij}^A + m_{B\to A}\frac{N_B}{N_A}\frac{\mathcal{Z}_A}{\mathcal{Z}_B}
 \left( J_{ij}^A - J_{ij}^B \right).
\label{eq:fij-inference-formula-v1_app_1}
\end{equation}
If in fact the relative sizes of the populations $A$ and $B$ have equilibrated
we have 
$N_B/N_A=m_{A\to B}/m_{B\to A}$. The inference formula becomes
\begin{equation}
f_{ij}^{A\note{,*}} =  (4\mu +r c_{ij}) J_{ij}^A + m_{A\to B}\frac{\mathcal{Z}_A}{\mathcal{Z}_B}
 \left( J_{ij}^A - J_{ij}^B \right).
\label{eq:fij-inference-formula-v1_app}
\end{equation}
Inserting the mean-field expression for the ratio of partition functions derived in the following Section~\ref{sec:inference-mean-field}
and given in \eqref{eq:fraction_partition_app}
we have a fully explicit inference formula in terms of characteristics of the data and the other parameters ($r,\mu,m_{A\to B},m_{B\to A}$) which have to be assumed known by other means.  

The equation 
\eqref{eq:additive_fitness_of_popA} for $h_i$ is on the other hand not of the relaxation type. It can be expected that either the allele at locus $i$ is fixated, in which case $h_i$ is practically $\pm\infty$, does not change, and there is no variability from which to infer fitness differences, or the frequency of alleles at locus $i$ changes in time.
This then leads to an inference formula with $\dot{h}_i$ on the right hand side:
\begin{equation}
{f}_i^{A,*} = \dot{h}_i^A+ 2\mu h_i^A - r\sum_{j}c_{ij} J_{ij}^A \chi_j^A + 
m_{B\to A}
\frac{N_B}{N_A} \frac{\mathcal{Z}_A}{\mathcal{Z}_B} (h_i^A - h_i^B)
\label{eq:fi-inference-formula-v1_app}
\end{equation}
Here $J_{ij}^A$ can be taken to be at its 
equilibrium value given by \eqref{eq:matrix_epistatis_eq}, and the ratio of the partition functions is given by \eqref{eq:fraction_partition_app}
which makes it a function of the magnetizations (one-locus allele averages) $\chi_i^A,\chi_i^B$
and $J_{ij}^A,J_{ij}^B$ and the overall parameters.
Again, if in fact the relative sizes of the populations $A$ and $B$ have equilibrated
we have 
$N_B/N_A=m_{A\to B}/m_{B\to A}$ which simplifies this second inference relation to 
\begin{equation}
{f}_i^{A,*} = \dot{h}_i^A +2\mu h_i^A - r\sum_{j}c_{ij} J_{ij}^A \chi_j^A + 2\mu h_i^A + m_{A\to B}  \frac{\mathcal{Z}_A}{\mathcal{Z}_B} (h_i^A - h_i^B)
\label{eq:additive_fitness_of_popA_eq}
\end{equation}
Equation~\eqref{eq:additive_fitness_of_popA_eq} can only be used for inference over relatively short time intervals such that 
$h_i^A$,
$\frac{\mathcal{Z}_A}{\mathcal{Z}_B}$ and $(h_i^A - h_i^B)$ stay roughly constant. However, this does not mean that the term $\dot{h}_i^A$ can be neglected; in general it will be of similar size as three other terms on the right hand side. 
\subsubsection{Note on the expression of the pairwise coupling in the case of symmetric migration}
Going back to the expression of the pairwise coupling:
\begin{equation}
\label{eq:inference_epistatis_mig_app}
    J_{ij}^A = \frac{1}{2(4\mu +r c_{ij})} \left[ 
    \left(f_{ij}^A + f_{ij}^B\right) 
    + \left(f_{ij}^A - f_{ij}^B\right)\frac{ 
    4\mu +r c_{ij} + \tilde{m}_{B\to A} - \tilde{m}_{A\to B}}
    {4\mu +r c_{ij} +  \tilde{m}_{B\to A} +\tilde{m}_{A\to B}} 
    \right].
\end{equation}
In the case of symmetric migration $m_{B\to A}=m_{A\to B}=m$, it becomes:
\begin{equation*}
J_{ij}^A = \frac{1}{(4\mu+ rc_{ij})(4\mu+ rc_{ij}+2m)}\left[(4\mu+ rc_{ij}+m)f_{ij}^A+mf_{ij}^B\right].
\label{eq:inference_epistatis_sym_mig}
\end{equation*}
the pairwise coupling appears as the weighted average of the two epistatic fitnesses, with weights determined by the migration rates relative to the other evolutionary forces. When $m=0$, we get back the initial KNS inference formula, and when the effect of migration is dominant ($m\gg 4\mu + rc_{ij}$), the pairwise coupling reduces to the KNS inference formula applied to the mean epistatic fitness of the two subpopulations:
\begin{equation*}
J_{ij}^A = \frac{f_{ij}^A+f_{ij}^B}{2(4\mu+ rc_{ij})}.
\label{eq:inference_epistatis_sym_mig_app}
\end{equation*}
\subsubsection{Note on the expression of the local field when its derivative is zero}
Combining \eqref{eq:additive_fitness_of_popA_eq} and its equivalent for population B and assuming $\dot{\mathbf{h}}=0$,we have

\begin{equation}
h_i^A = \frac{1}{4\mu}\left[\left(\tilde{f}_i^A+\tilde{f}_i^B\right) + \left(\tilde{f}_i^A-\tilde{f}_i^B\right)\frac{2\mu + \tilde{m}_{B\to A} - \tilde{m}_{A\to B}}{2\mu + \tilde{m}_{B\to A} + \tilde{m}_{A\to B}}\right]
\label{eq:inference_additive_mig}
\end{equation}
with $\tilde{m}_{A\to B}=m_{A\to B}\frac{\mathcal{Z}_A}{\mathcal{Z}_B}$  and $\tilde{f}_i^A=f_i^A+r\sum_{j\neq i}c_{ij}J_{ij}^A\chi_j^A$ the effective additive fitness due to the effects of recombination, and similarly for $\tilde{m}_{B\to A}$ and $\tilde{f}_i^B$. 
We note that if $\tilde{f}_{i}^A=\tilde{f}_{i} ^B=\tilde{f}_{i}$, then
\begin{equation*}
h_i^A = \frac{\tilde{f}_{i}}{2\mu} = h_{i} = h_{i}^B.
\label{eq:J_{ij}^A = J_{ij}^B_app}
\end{equation*}
Effectively the two populations then act, for the purpose of inference of additive fitness, as one larger combined population. If there is no migration, $\tilde{m}_{A\to B}=\tilde{m}_{B\to A}=0$, we also find the initial equilibrium value of $h_i^A=\tilde{f}_{i}^A/2\mu$, which was to be expected.\\

\subsection{Inference formulas in the Mean Field ansatz}
\label{sec:inference-mean-field}
The goal of this section is to
express the fraction $\mathcal{Z}_A/\mathcal{Z}_B$ using a mean field approximation. 
In this class of approximation the probability
$P_{A}(g)$ is approximated by a probability
distribution which is factorized over loci
$P_{A}^{fact}(g)=\prod_i P_{A,i}(s_i)$.
In the approximation known as 
"naive mean-field"
\cite{Tanaka2000,ricci2012bethe,Amari2016,Nguyen2017} the approximation is through the free energy
functional which is
\begin{equation}
    \label{eq:free_energy}
F_{nMF} = -\sum_{i=1}^{L}\left[H\left(\frac{1+\chi_i}{2}\right) +H\left(\frac{1-\chi_i}{2}\right)\right]
- \sum_i h_i \chi_i - \sum_{i<j} J_{ij} \chi_i \chi_j,
\end{equation}
where $H(x)=-x\log(x)$ and the $\chi_i$ are the magnetizations defined above.
The first sum in the expression of the free energy is the entropy of a factorized probability distribution and the second is the expected energy. Mean field approximations beyond naive mean field can be understood as making more advanced approximations of the entropy term~\cite{ricci2012bethe,Nguyen2017}.  
On the naive mean field level the 
magnetizations are solutions to of self-consistency equations:
\begin{equation}
    \label{eq:magnetizations}
\chi_i = \tanh \left(h_i +\sum_{\substack{j=1 \\ j \neq i}}^L J_{ij} \chi_j\right).
\end{equation}
We write \(\chi_i^{B}=\chi_i^{A}+\delta\chi_i\), 
and express the entropy term for subpopulation $B$ as
\begin{align}
\label{eq:H-H}
H\left(\frac{1+\chi_i^{B}}{2}\right) +H\left(\frac{1-\chi_i^{B}}{2}\right) 
=& H\left(\frac{1+\chi_i^{A}}{2}\right) +H\left(\frac{1-\chi_i^{A}}{2}\right) \notag \\
&-\frac{1}{2}\log\left(\frac{1-{\chi_i^B}^2}{1-{\chi_i^A}^2}\right)+\frac{\chi_i^B}{2}\log\left(\frac{1-\chi_i^B}{1+\chi_i^B}\right)-\frac{\chi_i^A}{2}\log\left(\frac{1-\chi_i^A}{1+\chi_i^A}\right).
\end{align}
The derivation is presented separately below in
Section~\ref{sec:entropy-term-difference}.
Using the self-consistency equations for the magnetizations eq.\eqref{eq:magnetizations} and the identity \(\log\left({\frac{1+x}{1-x}}\right)=2 \arctanh x\)
we can can re-write the
last two terms in
\eqref{eq:H-H}
as expressions of the type
\begin{equation*}
    \frac{\chi_i}{2}\log\left(\frac{1 - \chi_i}{1 +\chi_i} \right) =-\chi_i \arctanh \chi_i =-\left(h_i\chi_i  +  \sum_{\substack{j=1 \\ j \neq i}}^L J_{ij} \chi_i\chi_j\right).
\end{equation*}
Inserting this in \eqref{eq:H-H} we get
\begin{align*}
H\left(\frac{1+\chi_i^{B}}{2}\right) +H\left(\frac{1-\chi_i^{B}}{2}\right)=&H\left(\frac{1+\chi_i^{A}}{2}\right) +H\left(\frac{1-\chi_i^{A}}{2}\right) 
-\frac{1}{2}\log\left(\frac{1-{\chi_i^B}^2}{1-{\chi_i^A}^2}\right)\\ 
&+\left(-h_i^B\chi_i^B - \sum_{\substack{j=1 \\ j \neq i}}^L J_{ij}^B \chi_i^B\chi_j^B+h_i^A\chi_i^A + \sum_{\substack{j=1 \\ j \neq i}}^L J_{ij}^A \chi_i^A\chi_j^A\right).
\end{align*}

Then, using $\sum_{i<j} = \frac{1}{2} \sum_{i \neq j}$ with symmetric $J_{ij}$:
\begin{align*}
F_B-F_A=& \sum_{i=1}^L\frac{1}{2}\log\left(\frac{1-{\chi_i^B}^2}{1-{\chi_i^A}^2}\right)
            - \sum_{i=1}^L \left( h_i^A\chi_i^A -h_i^B\chi_i^B +h_i^B\chi_i^B-h_i^A\chi_i^A\right)\\
            & \quad \;\;
            - \sum_{i \neq j} \left[
            J_{ij}^A \chi_i^A\chi_j^A
           -J_{ij}^B \chi_i^B\chi_j^B
           + \frac{1}{2} \left(J_{ij}^B \chi_i^B\chi_j^B
           -J_{ij}^A \chi_i^A\chi_j^A\right) \right]\\
           &=\frac{1}{2}\sum_{i=1}^L\log\left(\frac{1-{\chi_i^B}^2}{1-{\chi_i^A}^2}\right) + \frac{1}{2}  \sum_{i \neq j} J_{ij}^B \chi_i^B\chi_j^B - J_{ij}^A \chi_i^A\chi_j^A.
\end{align*}
Using the definition $F=-\log(\mathcal{Z})$, the left hand side is the same as
$\log{\left(\mathcal{Z}_A/\mathcal{Z}_B\right)}$. We thus have:
\begin{align}
\frac{\mathcal{Z}_A}{\mathcal{Z}_B} 
&= \exp\!\left(\frac{1}{2}\left[
\sum_{i=1}^L \log\frac{1-{\chi_i^B}^2}{1-{\chi_i^A}^2}
+ {\boldsymbol{\chi}^B}^\top J^B \boldsymbol{\chi}^B 
- {\boldsymbol{\chi}^A}^\top J^A \boldsymbol{\chi}^A
\right]\right) \notag\\[6pt]
&= \prod_{i=1}^L \sqrt{\frac{1-{\chi_i^B}^2}{1-{\chi_i^A}^2}}
\cdot \exp\!\left(\frac{1}{2}\left[
{\boldsymbol{\chi}^B}^\top J^B \boldsymbol{\chi}^B 
- {\boldsymbol{\chi}^A}^\top J^A \boldsymbol{\chi}^A
\right]\right).
\label{eq:fraction_partition_app}
\end{align}

The first term of this equation is due to the differences in one-locus magnetizations  between the two systems, while the exponential comes from the pairwise interactions. As required for the consistency of this equation, we find the inverse of this ratio by inverting the indices $A \leftrightarrow B$. 
\subsubsection{Entropy terms difference in the ratio of partition functions}
\label{sec:entropy-term-difference}
Again, we write \(\chi_i^{B}=\chi_i^{A}+\delta\chi_i\), then
\begin{equation}
\begin{aligned}
H\left(\frac{1+\chi_i^{B}}{2}\right)&=-\left(\frac{1+\chi_i^{A}+\delta\chi_i}{2}\right)\left[\log\left(\frac{1+\chi_i^{A}}{2}\right)+\log\left(1+\frac{\delta\chi_i}{1+\chi_i^{A}}\right)\right]\\
&=-\left(\frac{1+\chi_i^{A}+\delta\chi_i}{2}\right)\left[\log\left(\frac{1+\chi_i^{A}}{2}\right)+\log\left(\frac{1+\chi_i^B}{1+\chi_i^{A}}\right)\right]\\
&=H\left(\frac{1+\chi_i^{A}}{2}\right)-\frac{\delta\chi_i}{2}\log\left(\frac{1+\chi_i^{A}}{2}\right)-\frac{1}{2}\log\left(\frac{1+\chi_i^{B}}{1+\chi_i^{A}}\right)-\frac{\chi_i^B}{2}\log\left(\frac{1+\chi_i^{B}}{1+\chi_i^{A}}\right)\\
\end{aligned}
\end{equation}
The same goes for the \((1-\chi_i^{B})/2\) term:
\begin{equation}
\begin{aligned}
H\left(\frac{1-\chi_i^{B}}{2}\right)=&H\left(\frac{1-\chi_i^{A}}{2}\right)+\frac{\delta\chi_i}{2}\log\left(\frac{1-\chi_i^{A}}{2}\right)-\frac{1}{2}\log\left(\frac{1-\chi_i^{B}}{1-\chi_i^{A}}\right)+\frac{\chi_i^B}{2}\log\left(\frac{1-\chi_i^{B}}{1-\chi_i^{A}}\right)\\
\end{aligned}
\end{equation}
Summing the two terms, we get
\begin{equation}
\begin{aligned} 
H\left(\frac{1+\chi_i^{B}}{2}\right) + H\left(\frac{1-\chi_i^{B}}{2}\right) =& H\left(\frac{1+\chi_i^{A}}{2}\right) + H\left(\frac{1-\chi_i^{A}}{2}\right)- \frac{1}{2} \log\left(\frac{1-{\chi_i^{B}}^2}{1-{\chi_i^{A}}^2}\right) \\  
&\quad\;+ \frac{\chi_i^B}{2} \log\left(\frac{1-\chi_i^{B}}{1-\chi_i^{A}}\right) - \frac{\chi_i^B}{2} \log\left(\frac{1+\chi_i^{B}}{1+\chi_i^{A}}\right) + \frac{\delta\chi_i}{2} \log\left(\frac{1-\chi_i^{A}}{1+\chi_i^{A}}\right)\\[8pt]
=&H\left(\frac{1+\chi_i^{A}}{2}\right) + H\left(\frac{1-\chi_i^{A}}{2}\right) \\
&\quad\;-\frac{1}{2} \log\left(\frac{1-{\chi_i^{B}}^2}{1-{\chi_i^{A}}^2}\right)+\frac{\chi_i^B}{2} \log\left(\frac{1-\chi_i^{B}}{1+\chi_i^{B}}\right)-\frac{\chi_i^A}{2} \log\left(\frac{1-\chi_i^{A}}{1+\chi_i^{A}}\right).
\end{aligned}
\end{equation}
which is \eqref{eq:H-H} above.

\subsection{Supporting results}
\begin{figure}[ht!] 
\centering
\includegraphics[width=0.4\linewidth]{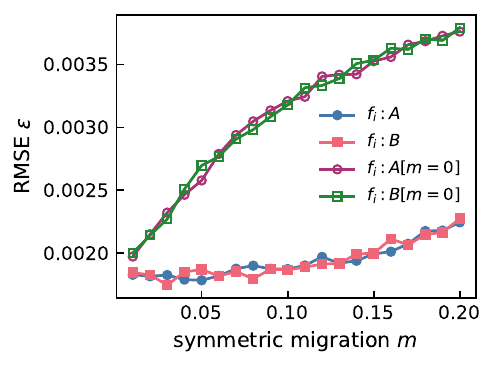}
\includegraphics[width=0.4\linewidth]{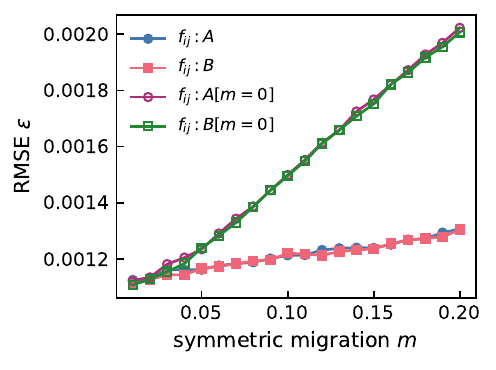}\\
\includegraphics[width=0.4\linewidth]
{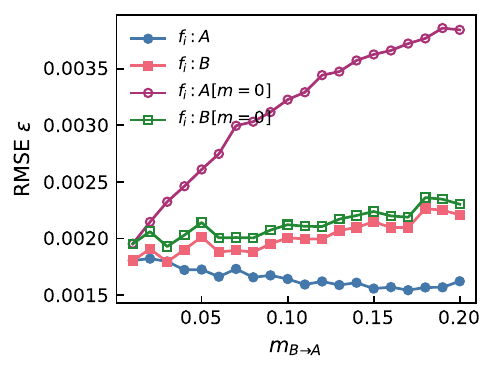}
\includegraphics[width=0.4\linewidth]{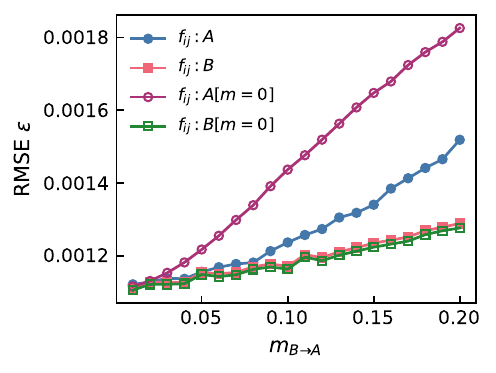}
\caption{
Supplementary analysis of the effects of migration on fitness parameter inference when the additive fitness terms of subpopulations $A$ and $B$ are sampled from the same Gaussian distribution with identical mean and variance.  
The RMSE $\varepsilon$ is shown as a function of the migration rate for additive fitness terms $f_i$ and epistatic couplings $f_{ij}$. 
Panels (a) and (b) show the results under symmetric migration, with $m_{A\to B}=m_{B\to A}=m$. 
Panels (c) and (d) show the corresponding results under asymmetric migration, where $m_{A\to B}=0.01$ is fixed and $m_{B\to A}$ is varied. 
Filled symbols denote results obtained using the migration inference theory, whereas open symbols denote the no-migration approximation in which migration terms are neglected by setting $m=0$. 
The overall trends are consistent with those observed in the main text: accounting for migration term reduces the inference error for both additive and epistatic fitness parameters, while asymmetric migration leads to more population-dependent behavior.
}
\label{fig:supp_same_fitness_distribution}
\end{figure}
\subsubsection{Effects of migration 
on the inference on fitness parameters}

As a supplementary test, we examined the case in which populations $A$ and $B$ are sampled from the same underlying Gaussian fitness distribution, with identical mean and variance. 
This differs from the main-text setting, where the two subpopulations have different mean additive fitness coefficients. 
As shown in Fig.~\ref{fig:supp_same_fitness_distribution}, the resulting trends are qualitatively similar to those reported in the main text. 
Under symmetric migration, the migration-aware inference gives lower RMSEs than the no-migration approximation for both additive fitness terms $f_i$ and epistatic couplings $f_{ij}$. 
Under asymmetric migration, the inference error becomes more population dependent, but explicitly accounting for migration still generally improves or gives comparable reconstruction accuracy relative to the no-migration approximation.

These results indicate that the migration induced changes in inference accuracy do not rely solely on intrinsic differences between the two fitness landscapes. 
Even when $A$ and $B$ are generated from the same Gaussian distribution, migration can still modify the allele-frequency and correlation structures used for inference. 
Therefore, the supplementary results support the robustness of the main conclusion that migration should be explicitly incorporated when inferring additive and epistatic fitness parameters in structured populations.

\begin{figure}[ht!]
    \centering
    \includegraphics[width=0.4\linewidth]{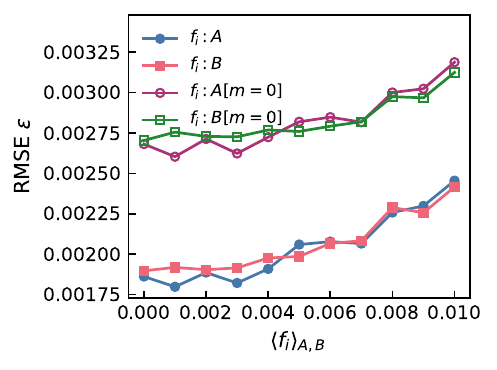}
    \includegraphics[width=0.4\linewidth]{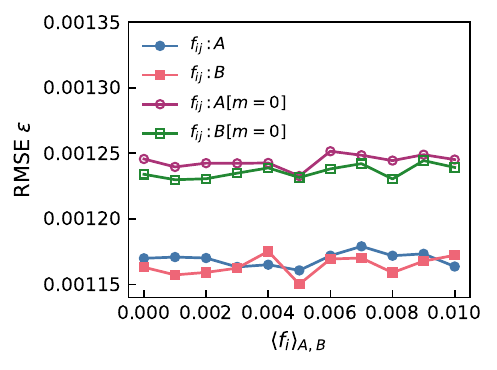}\\
    \includegraphics[width=0.4\linewidth]{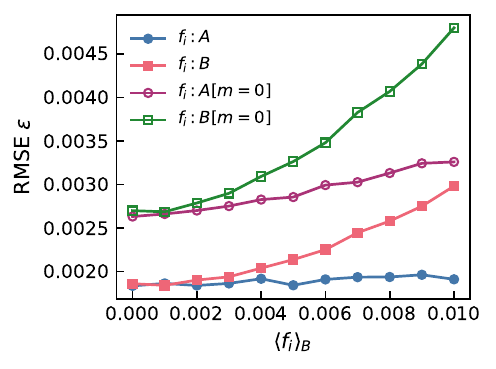}
    \includegraphics[width=0.4\linewidth]{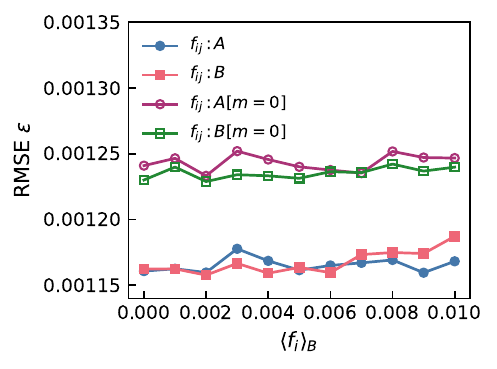}
    \caption{
Supplementary analysis of the effect of selection strength on fitness parameter inference under symmetric migration $m_{A\to B}=m_{B\to A}=0.05$. 
Panels (a) and (b) show the RMSE $\varepsilon$ of the inferred additive fitness terms $f_i$ and epistatic couplings $f_{ij}$, respectively, when the mean additive fitness coefficients of $A$ and $B$ are varied simultaneously, denoted by $\langle f_i\rangle_{A,B}$. 
Panels (c) and (d) show the corresponding results when only the mean additive fitness coefficient of $B$, $\langle f_i\rangle_B$ is varied, while that of $A$ is kept fixed at $\langle f_i\rangle_A=0$. 
Solids denote results obtained using the migration inference theory, whereas empties denote the no-migration approximation. 
The results show that increasing selection strength mainly increases the inference error of additive fitness terms, particularly in the population whose selection strength is varied. 
In contrast, the inference error of epistatic couplings remains nearly insensitive to the mean additive fitness coefficient. 
The migration theory generally gives lower RMSEs than the no-migration approximation for both additive and epistatic fitness inference.
}
\label{fig:supp_selection_symmetric}
\end{figure}

\subsubsection{Effect of standard deviation of additive fitness}
We further examined the effect of selection strength on fitness parameter inference under symmetric migration with $m_{A\to B}=m_{B\to A}=0.05$. 
As shown in Fig.~\ref{fig:supp_selection_symmetric}, increasing the mean additive fitness of two subpopulations  mainly affects the inference of additive fitness terms $f_i$. 
When $\langle f_i\rangle_{A,B}$ is varied simultaneously, the RMSE of $f_i$ increases for both populations. 
When only $\langle f_i\rangle_B$ is varied, the increase in RMSE is mainly observed in $B$, whereas the error in $A$ remains relatively stable. 
In contrast, the inference error of epistatic couplings $f_{ij}$ is nearly insensitive to the variation of the mean additive fitness coefficient in both settings. 
Compared with the no-migration approximation, the migration theory generally yields lower RMSEs, for both additive and epistatic fitness inference. 
These results further support the main conclusion that selection strength primarily affects additive fitness reconstruction, whereas epistatic inference is less sensitive to $\langle f_i\rangle$ under the parameter regime considered here.

\bibliography{refs}  
\end{document}